\documentclass[12pt]{iopart}

\usepackage{graphicx}
\usepackage{longtable}
\usepackage{subfigure}
\renewcommand{\contentsname}{}

\begin{document}
\title[Interaction of eta mesons with nuclei]{Interaction of eta mesons with nuclei} 

\author{
N G Kelkar$^{1,\dagger}$, K  P Khemchandani$^{2, \dagger\dagger}$, 
N J Upadhyay$^{3, ^*}$ and
B K Jain$^{4, ^{**}}$}
\address{$^1$ Dept. de Fisica, Univ. de los Andes, Cra 1E, 18A-10, Bogota, Colombia}
\address{$^2$ Instituto de F\'isica, Universidade de S\~ao Paulo, C.P. 66318, 05389-970 S\~ao 
Paulo, SP, Brazil}
\address{$^3$ National Superconducting Cyclotron Laboratory, 
Michigan State University, East Lansing, Michigan 48824-1321, USA}
\address{$^4$ MU-DAE Centre for Excellence in Basic Sciences, Mumbai University,
Mumbai 400098, India}
\ead{$^\dagger$nkelkar@uniandes.edu.co, $^{\dagger \dagger}$kanchan@if.usp.br, 
$^*$upadhyay@nscl.msu.edu, $^{**}$brajeshk@gmail.com}
\begin{abstract}
Back in the mid eighties, a new branch of investigation which was related 
to the interaction of eta mesons with nuclei came into existence. It started 
with the theoretical prediction of possible exotic states of eta mesons and 
nuclei bound by the strong interaction and later developed into an 
extensive experimental program to search for such unstable states as well as 
understand the underlying interaction via eta meson producing reactions. 
The vast literature of experimental as well as theoretical works which 
studied various aspects of eta producing reactions such as the 
$\pi ^+$ $n$ 
$\to \eta p$, $p d \to ^3$He $\eta$, $p \,^6$Li $\to ^7$Be $\eta$ and 
$\gamma ^3$He $\to \eta$ X, to name a few, had but one objective in mind: 
to understand the eta - nucleon ($\eta N$) and hence the $\eta$-nucleus 
interaction which could explain the production data and confirm the 
existence of some $\eta$-mesic nuclei. In spite of these efforts, there remain 
uncertainties in the knowledge of the $\eta N$ and hence 
the $\eta$-nucleus interaction. 
The present work is hence an attempt to 
bind together the findings in these works and draw some global and specific conclusions which 
can be useful for future explorations.
 
The $\eta N$ scattering length (which represents the strength of the $\eta$ - 
nucleon interaction) using different theoretical models and analysing the 
data on $\eta$ production in pion, photon and proton induced reactions was 
found to be spread out in a wide range, namely, 
0.18 $\le$ $\Re$e a$_{\eta N}$ $\le$ 1.03 fm
and 0.16 $\le$ $\Im$m a$_{\eta N}$ $\le$ 0.49 fm.
Theoretical searches of heavy $\eta$-mesic nuclei based on $\eta$-nucleus 
optical potentials and lighter ones based on Faddeev type few body approaches
predict the existence of several quasibound and resonant states. Though 
some hints of $\eta$-mesic states such as $^3_{\eta}$He and $^{25}_{\eta}$Mg 
do exist from previous experiments, the promise of clearer signals for the 
existence of $\eta$-mesic nuclei lies in the experiments to be performed at 
the J-PARC, MAMI and COSY facilities in the near future. This review is 
aimed at giving an overall status of these efforts. 
\end{abstract}

\submitto{\RPP}
\maketitle
{\bf Contents}
\tableofcontents

\section{Introduction} 
Meson related physics has always led to interesting findings in 
different ways. Though mesons are strongly interacting objects composed of 
quark - antiquark ($q \bar{q}$) pairs, the bonding of electrically charged 
mesons with nuclei leads to the formation of exotic atoms. For example, 
negatively charged pions or kaons could replace an electron in an outer 
orbital in a standard atom and get bound in the atom due to the Coulomb 
interaction. After some transitions to lower states however, 
the meson comes within the range of the strong nuclear interaction and 
is absorbed on to the nucleus or lost in a nuclear reaction. Thus a meson 
which in principle in free space could have undergone a weak decay gets 
bound by the electromagnetic interaction and vanishes by the 
strong interaction. There exists yet another possibility for the formation of 
bound states of mesons and nuclei and this is when the meson-nucleus strong 
interaction is attractive (which may not necessarily be the case). 
The eta ($\eta$) meson seems to satisfy 
this requirement. Its interaction with the 
nucleon in the s-wave (which proceeds through the formation of an 
N$^*$(1535) nucleon resonance) was found to be attractive \cite{bhalerao} 
and led to the prediction \cite{firstheor1,firstheor2,firstheor3} 
and early searches \cite{firstexp} of unstable states of eta 
mesons and nuclei. The existence of unstable bound 
states of the strange $K$-meson and nuclei was also indicated in some 
recent experiments as a result of the $K^- p$ $s$-wave interaction
\cite{kmeson1, kmeson2}. Additionally, the superposition of a large 
repulsive s-wave $\pi ^-$-nucleus
interaction at low pion momenta and an attractive Coulomb interaction has 
been seen to give rise to deeply bound pion-nucleus
bound states \cite{bpion}.

The interaction of the $\eta$-meson with a nucleon near threshold is 
mainly determined by the $S_{11}$, $ J^{\pi}({\rm spin^{parity}}) 
= \frac {1}{2}^-$ 
resonance N$^*$(1535), which is just 49 MeV above the $\eta N$ threshold 
and has a width $\Gamma $=150 MeV, thus covering the whole low energy region 
of the $\eta N$ interaction. As the $S_{11}$-resonance also decays to 
$\pi N$, $\gamma N$ and $\pi \pi N$ channels, the correct treatment of the 
$\eta N$ interaction, therefore, involves its coupling to all these channels. 
Several such coupled channel calculations have been reported 
in literature \cite{etaN0,etaN00,etaN1,etaN2,etaN3,etaN4,etaN5}. 
Results of these calculations are fitted to the available data, and the 
$\eta N$ scattering amplitude is obtained. 
However, as the fitted data do not include the elastic $\eta N$ 
scattering data due to nonavailability
of $\eta$ beams, such extracted $\eta N$ scattering amplitudes and 
hence the $\eta N$ scattering lengths, $a_{\eta N}$,  
have uncertainties. But one conclusion which seems definite in all the 
calculations is that the $\eta N$ interaction is strong and attractive 
in the s-wave. 

When viewed as an elementary particle, the $\eta$-meson 
is quite similar to  the $\pi^0$-meson, despite the $\eta$ 
being about four times heavier than the $\pi^0$. 
Both mesons are pseudoscalar, i.e., 
$J^{\pi} = 0^{-}$, are charge neutral, 
have almost the same lifetime (~10$^{-18}$ sec) 
and are the only mesons that have a high probability of pure radiative decay. 
Structurally though, as they are the quantum states of $q\bar {q}$, 
it is believed that due to the mass difference between the u and d quarks 
and their electromagnetic interaction, 
the $\pi ^0$ and $\eta$ are not pure isotopic spin states. 
The physically observed $\pi^0$ and $\eta$ are, actually, 
the superposition of the pure isotopic spin T=0 and 1 quantum states. 
The value of the mixing angle of these two states obtained in 
\cite{abdel} from an analysis of the cross sections for the 
$p d \to \, ^3$H $\pi^+$/$^3$He $\pi^0$ reactions is 0.006 $\pm$0.005 radians. 
Ref. \cite{abdel} also discusses literature where the values vary between 
0.01 to 0.034 radians. 
The $\eta$-meson 
is also known to mix with its heavier partner, $\eta \prime $.
The isospin mixing structure of the $\pi^0$ and $\eta$ gets 
manifested as isospin symmetry breaking (ISB) and charge symmetry 
breaking (CSB) when these mesons come in contact with nucleons. 
Though it is believed that most of the ISB and CSB effects observed 
in hadronic systems have the above origin, significant and 
directly observable isospin-violating effects may be seen 
in pion production at energies near the $\eta$ threshold 
via $\eta-\pi^0$ external mixing. Such exploratory studies have been 
conducted experimentally and theoretically \cite{mix1,mix2,mix3}.

Coming back to the $\eta N$ interaction, its strong attractive nature
opened up the possibility of the formation of $\eta$-nucleus quasibound 
states. The initial theoretical prediction around 1986 
about the possible existence of 
unstable bound states of eta mesons and nuclei gave rise to 
experimental programs to hunt for the existence of eta-mesic states as well 
as eta meson production experiments aimed at studying the eta-nucleus 
interaction in the final states. The theoretical developments kept
pace with the experiments. Over the years more and more sophisticated 
models to understand the basic eta-nucleon interaction, 
the few body problem of eta mesons and light nuclei and
the many body problem of eta mesons and heavy nuclei kept appearing. After 
about 25 years of investigations in this field, 
there has been a lot of progress and some evidence exists 
for the existence of such 
states. However, there is still no general agreement and no 
final word on the strength of the eta-nucleon and 
eta-nucleus interaction.

Though a detailed account of the various theoretical and experimental efforts 
will be given in the next sections of the article, at this point we mention 
a few of these works. 
For nuclei beyond few-nucleon systems, calculations have mostly been done 
using an $\eta$-nucleus optical potential. 
In \cite{firstheor1,firstheor2,opticalhaider}, the $\eta$-mesic states were 
found from complex energy solutions 
of the momentum-space relativistic three-dimensional integral equation using 
different optical potentials. 
In this work, the binding energies and widths were found to strongly 
depend on the sub-threshold $\eta$-nucleon interaction. 
These calculations predicted unstable bound or 
quasibound states for nuclei with mass number 
greater than 10. In another approach using optical potentials  
\cite{opticaloset1}, the self energies of the $\eta$ meson 
in the nuclear medium evaluated for 
$^{12}$C, $^{40}$Ca and $^{208}$Pb were used to find the 
widths of quasibound states of $\eta$ mesons and nuclei. In 
\cite{opticaloset2}, the energies and widths of 
several $\eta$-mesic nuclei 
were predicted by once again using self energies of the $\eta$ in the medium 
but evaluated using the technique of unitarized chiral perturbation theory.
Another approach that has been used is the QCD based 
quark-meson-coupling approach (QMC), where the $\eta$-meson is 
embedded in the nuclear medium and couples to quarks and 
mixes with $\eta \prime $ \cite{opticalthomas1,opticalthomas2}. 
The $\eta$ self energy obtained 
from such calculations was used in the local density approximation  
in the Klein-Gordon equation to obtain the complex energy solutions. 
This approach, apart from predicting the energies and widths  
of the quasibound states also showed the important role of 
$\eta - \eta \prime $ mixing in understanding the $\eta N$ 
scattering amplitude.

For few-nucleon systems, the existence of $\eta$ binding is mostly explored by 
calculating the poles in the scattering amplitude and the 
corresponding $\eta$-nucleus scattering lengths. 
The existence of quasibound states requires the real parts of the 
scattering lengths to be large and negative. 
In \cite{ueda} the author predicted a quasibound state in the 
$\eta N N$-$\pi N N$  
coupled system and confirmed its existence through the existence of a pole 
and a remarkable enhancement of the $\eta d$ elastic cross section. The state 
was predicted with a mass of 2430 MeV and a width of 10-20 MeV. 
In \cite{garci1,garci2,garci3} using Faddeev equations and a certain 
choice of potentials for the binary sub-systems in it, 
the authors found at best the existence of a quasivirtual 
state only. Other works involve the use of a multiple scattering 
formalism \cite{wycech1,wycech2} and calculations done within the finite
rank approximation using few body equations \cite{rakit1,rakit2,rakit3} 
to characterize $\eta$-mesic states of the deuteron, $^3$H, 
$^3$He and $^4$He nuclei. These 
and other few body calculations will be discussed in detail in section 3.

The anticipated $\eta $-nucleus bound states should get 
reflected in a straight forward way in the measured eta producing nuclear 
cross sections. Such observations have indeed been made in the proton 
induced $\eta$ production reactions where the measured amplitude 
on the deuteron target shows a sharp rise as one approaches the 
threshold \cite{dataMayer,dataBerger,dataWronska,dataMersmann,dataRausmann} 
of $\eta$ production. 
The $\eta$ $^3$He (or $\eta d$) scattering 
amplitude has been usually extracted from data 
by writing the cross section 
as a product of the plane wave result and an enhancement factor, which 
has been parametrized in terms of the scattering length \cite{parascatt}. 
Large values of the real part of the 
extracted scattering lengths from such analyses 
are normally taken as an indicator for the existence of bound states. 
This procedure of extracting the scattering lengths, 
however, ignores reference to any reaction mechanism for $\eta$ 
production and lacks a detailed treatment of the final state 
interaction. Therefore, while the extracted scattering lengths by this 
method may at best be indicative, their actual values may not be 
fully reliable for determining the bound states. 
A proper procedure should be to first analyze these data using a certain 
$\eta$ production mechanism \cite{lagetmechanisms} 
and then incorporate in detail the 
final state interaction (FSI) as accurately as possible. 
Once such a procedure, using the available $\eta N$ transition matrix, 
reproduces the measured cross sections, 
one can then infer from them the resultant $\eta$-nucleus 
scattering amplitudes, which will have a certain in-built off-shell behaviour. 
These amplitudes can then be used to investigate the 
 possibility of the existence of unstable bound states.  
Based on this concept the present authors first 
carried out a systematic analysis of the $\eta$ producing nuclear 
reactions to obtain a reliable $\eta$-nucleus scattering amplitude  
which was then used to locate the unstable bound states in the 
$\eta$-deuteron, $\eta$-$^3$He and $\eta$-$^4$He systems \cite{ourtimedelay}.  
The states were located using Wigner's time delay method which had 
earlier been successfully used to locate meson and baryon resonances 
\cite{hadrontimedelay1,hadrontimedelay2}. 
A modification of Wigner's
method was done in \cite{meprl} and an eta-mesic state in the $\eta$-$^3$He
system was found in agreement with experiments.

As for the experimental status of eta mesic states, 
some measurements which give a positive indication of
the existence of $\eta$-mesic states 
have been reported in literature.
One such experiment was performed by the TAPS collaboration
\cite{taps} on the photo-production of $\eta$ on $^3$He, namely, 
$\gamma \,^3$He $\to
\, \pi^0 p$ X, where one essentially sees the decay of a bound
$\eta$ in $^3$He through the
$S_{11}$ resonance. The other one is a bit
more recent measurement from COSY on the
$p \,^{27}$Al $\to \,^3$He X reaction in a recoil free kinematic set-up
\cite{cosy, betigeri},
where one observes in coincidence with $^3$He,
the decay of a possible bound $\eta$-$^{25}$Mg  state,
again, through the $S_{11}$ resonance. A detailed account of the experimental 
status will be given in section 3. 

In addition to the above studies of eta-mesic quasibound states, 
large efforts have also gone  
in understanding the eta producing reactions which explore 
the $\eta$ production vertex in these reactions and the effect 
of the eta interaction with different nuclei in the final state. 
The elementary reaction $N N \to N N \eta$ has been studied 
extensively within boson exchange models which include the 
exchange of $\pi$, $\eta$, $\rho$ and $\omega$ mesons. The opinion
regarding the role of these mesons seems to be however divided. In one of the 
first works \cite{lagetppeta,wilkinppeta} and more recent 
ones \cite{santraBK1},  
the $\rho$ meson was found to play a dominant role. However, measurements of 
the analyzing power for the $\vec{{p}}$ $p \to p p \eta$ reaction 
(with $\vec{p}$ indicating a polarized proton beam) 
near threshold \cite{czyzykiewicz} and theoretical calculations 
in \cite{rshyamppeta} up to 10 GeV beam energy do not agree with a $\rho$ 
meson dominance. The production of $\eta$ mesons in other reactions such 
as the $p d \to$ $\, ^3$He $\eta$, $p d \to$ $p d \eta$ and 
$d d \to$ $\, ^4$He $\eta$ has been studied \cite{wilkin1,wilkin2,wilkin3, 
kondratyuk,santraBK2,kanchan1,kanchan2,kanchan3,neelam1,neelam2}
within one-, two- and 
three-body exchange mechanisms proposed in \cite{lagetmechanisms}. 
A detailed account of these works as well as those on the $p \,^6$Li $\to$ 
$\,^7$Be $\eta$ reaction \cite{khalili,neelam7Be} 
and the available measurements \cite{dataMayer,dataBerger,dataMersmann,
dataRausmann,parascatt, berth,pd3He2,
pd3He3,pd3He4,pd3He5,hib00, piskor,
scomp,ulc,mach,gemcosy} will 
be given in section 4. We note here in passing that 
though the three body mechanism where the large momentum transfer 
in the production of the massive $\eta$ meson is shared between nucleons 
seems to reproduce the threshold data very well, it fails to reproduce the 
forward peaks in the angular distributions at high energies and thus 
remains to be an open problem to be settled in future.

This report is organized as follows. 
As the $\eta N$ scattering amplitude plays a pivotal role in all the
eta-nucleus studies, we begin by giving a comprehensive account on the present 
knowledge of the elementary amplitude and the procedures employed to
obtain it. A large section after this is dedicated to present 
an exhaustive account of the experimental 
and theoretical searches for the existence of 
the exotic eta-mesic states. Considering the vast amount of literature 
available, though we will try to provide a complete bibliography, 
the focus of these sections will be on a few theoretical works representative 
of each type of formalism. In addition to the studies of eta mesic 
quasibound states, huge efforts have also gone in understanding the 
eta producing reactions which explore the eta production vertex in 
these reactions and the effect of the eta interaction with other nuclei in 
the final state. The reaction mechanisms used for eta production are 
usually based on models similar to those used for other mesons such as 
the pions and kaons. Hence after giving a brief description 
of all available mechanisms and their success in comparison with data, 
we shall review their relevance for the eta production reactions near 
threshold as well as at high energies in Section 4. 
Since the objective of the eta 
producing reactions is to finally investigate the eta nucleus interaction 
in the final state, the next section naturally goes over to discuss the 
results obtained as compared to data. 
Here, in Section 5, the focus will be on specific reactions, the theoretical 
estimates for them and their comparison with data.
Finally, after having surveyed most of the existing literature on 
theoretical predictions and experimental data, we will try to draw 
conclusions on the range of the strength of the eta nucleon and hence eta 
nucleus interaction. These conclusions should help in planning experiments 
for future investigations in this field.

\section{$\eta $-nucleon interaction and scattering amplitude}
The most appropriate source of the $\eta$-nucleon scattering amplitude 
should, of course, be  the elastic 
scattering data on it. However, such studies  can not be pursued because 
$\eta $ beams are not available. The next appropriate 
way then is to obtain it from   
the $\eta$ production reactions like $\pi N\rightarrow \eta N$ and 
$\gamma N\rightarrow \eta N$, where this information appears 
in the final state through the interaction of the emerging $\eta $-meson 
with the recoiling nucleon. Detailed experimental information on these 
reactions exists now as the pion and photon beams are 
 in use over a wide range of energies. The database includes the total and the differential cross sections from threshold to high energies. It also 
includes some polarization measurements.  The older of these data have been reviewed in \cite{cs, abs}, while the most recent ones are given 
in \cite{prakhov, abm}. Plotted as a function of beam energy the total 
pion-nucleon cross section shows a sharp growth 
above the $\eta $-threshold, signalling the opening of the $\eta N$ channel. 
The sharp rise in the pion-induced eta cross section 
is attributed to the broad nucleon resonance, 
$S_{11}(1535)$, which is very close to 
the $\eta $-production threshold at $\sqrt {s}$ $\simeq$ 1487 MeV. It is 
strongly coupled to both the pion and the eta-meson in the s-wave. 
The other resonances near the $\eta $-threshold are the $P_{11}(1440)$ 
and $D_{13}(1520)$. However, the coupling of $D_{13}(1520)$ to 
the $\eta N$ channel is very weak 
(branching ratio $\Gamma _{\eta N}(1520)/\Gamma _{tot}$=0.0023$\pm $0.0004)   
and that of the subthreshold but very broad $P_{11}(1440)$ 
undetermined. 
With increasing energy, the $D_{13}(1520)$ shows up through an interference 
with the dominant $S_{11}(1535)$.

Near threshold the $\eta$-nucleon scattering thus 
consists of elastic scattering, $\eta N\rightarrow \eta N$ and the reactive 
$\eta N\rightarrow \pi N$ channel. It also has another reactive channel 
$\eta N\rightarrow \pi \pi N$ (arising from the $\pi \Delta$ and 
$\rho N$ couplings), but the cross section for it is very small. 
The $\eta $-nucleon scattering length $a_{\eta N}$, which is the parametrization of the $\eta $-nucleon scattering amplitude at low energies, therefore, 
is complex. Its imaginary part gives a measure of the reactive content of the 
cross section. Since through the detailed balance theorem, 
the $\eta N\rightarrow  \pi N$ 
cross section can be related to the $\pi N\rightarrow \eta N$ cross section at an appropriate energy, the imaginary part of $a_{\eta N}$ 
can be determined directly by the pion-induced eta production data. In \cite{abm} in fact, using this and the optical theorem, a lower limit 
was set on the value 
of $\Im$m ($a_{\eta N}$) in the following way: The optical theorem gives 
\begin{equation}
\Im m (a_{\eta N})=\frac {q_\eta}{4\pi}\sigma _{\eta N}^{tot}=\frac {q_\eta}{4\pi}(\sigma _{\eta N\rightarrow \pi N}+
\sigma _{\eta N\rightarrow 2\pi N}+
\sigma _{\eta N\rightarrow \eta N}).
\end{equation}
Applying the detailed balance theorem  to it we get
\begin{equation}
\Im m (a_{\eta N})=\frac {3q_\pi ^2}{8\pi q_\eta }\sigma _{\pi ^-p\rightarrow \eta n}+
\frac {q_\eta }{4\pi }(\sigma _{\eta N\rightarrow 2\pi N}+
\sigma _{\eta N\rightarrow \eta N}),
\end{equation}
which results in
\begin{equation}
\Im m (a_{\eta N})\geq \frac {3q_\pi ^2}{8\pi q_\eta }\sigma _{\pi ^-p\rightarrow \eta n}.
\end{equation}
Using the recent threshold data \cite{prakhov}, $\sigma _{\pi ^-p\rightarrow \eta n}/q_\eta $=15.2$\pm$0.8 $\mu $b/MeV, gives
\begin{equation}
\Im m (a_{\eta N})\geq 0.172\pm0.009 \, {\rm fm}.
\end{equation}
In the above expressions, $q_x$ is the centre of mass 
momentum of the particle $x$.

With the real part of $a_{\eta N}$, however, this is not the case. 
The $\eta N$ channel in the $\pi N$ elastic scattering shows up  
as a cusp at the $\eta $ threshold, from where one can not directly 
determine the value of the real part of $a_{\eta N}$. 
The difficulty in obtaining Re($a_{\eta N}$) can also be seen from 
the behaviour of a resonance amplitude as a function of energy. 
A typical scattering amplitude proceeding through a resonance has a 
characteristic behaviour that, its 
real part goes through zero at the resonance position 
and the imaginary part peaks at the same position. 
Knowing that the $\eta N$ scattering 
at low energies is dominated by the $S_{11}(1535)$ resonance, 
the real part of the $\eta N$ amplitude 
would go through zero at resonance, making it 
difficult thereby to determine the precise value of the real part of 
$a_{\eta N}$ which receives contributions from non resonant processes. 

\subsection{Phenomenological}
The $\eta N$ scattering amplitude in literature has been obtained mostly 
phenomenologically by analysing the  
data on the $\eta $ producing  pion- and 
gamma-induced reactions, $(\pi , \eta )$ and $(\gamma , \eta )$. Since hadronic and electromagnetic $\eta $ production processes are 
different, details 
of $\eta N$ scattering  extracted from them complement each other. In different analyses, the four channels, $\gamma N$, $\pi N$, $\pi \pi N$, 
and $\eta N$ are included and  coupled-channels analyses are performed. In 
all calculations efforts are always made to include as broad a 
database as possible and available at the time of a particular calculation. 
Broadly, these efforts can be classified into two categories. 
In one class the available 
data are directly fitted to the T-matrices for the pion elastic scattering and pion- and gamma induced eta production. 
In another class, a microscopic model is developed to describe the reaction dynamics, and the parameters of this model are fixed by fitting the 
data. 
The $\eta N$ T-matrix is the outcome  
 of these calculations. These $\eta N$ T-matrices in the first approach are  only on-shell. 
For their off-shell application they need to be extrapolated 
with some ansatz. The second approach gives the  $\eta N$ T-matrix which has some inbuilt off-shell behaviour. 
However, because of the inherent fact that the information on the 
$\eta N$ T-matrix is an outcome of these calculations, the predicted 
values of $a_{\eta N}$ from all these calculations 
differ from one another. The real part of the $\eta N$ scattering length is 
found to have a large spread, from about 0.2 to 1 fm. 

The calculations in the first category are often done within the 
K-matrix approach, where the K-matrix is related to the usual T-matrix by 
the integral equation \cite{sauermann},
\begin{equation}
T=K-i\pi K\delta (E-H_0)T,
\end{equation}
where $H_0$ describes the free motion of the two interacting particles. The K-matrix is Hermitian, and the above transformation ensures 
that the T-matrix remains unitary.  Most detailed  
K-matrix analyses for the reactions of the 
present discussion have been carried out in \cite{etaN1, etaN2, abm}. 
It is gratifying to see that 
the latest calculations in \cite{etaN2} and \cite{abm} which use similar inputs give comparable values of $a_{\eta N}$ = 0.91 + $i$ 0.27 
and 1.14 + $i$ 0.31 fm, respectively. Furthermore, the extracted 
on-shell $\eta N$ T-matrix from these calculations is made useful to the 
few-body eta physics by first writing it in the ``effective range expansion" 
and then extrapolating it to the off-shell region by a 
simple separable approximation, i.e.
\begin{equation}\label{greenequation}
T_{\eta N}(q,E,q\prime)=v(q)t_{\eta N}(E)v(q\prime),
\end{equation}
with $v(q)=1/(1+q^2\beta ^2)$, where $\beta $ is the range parameter 
whose value is not well determined. 
In \cite{etaN2} it is taken = 0.31 fm. 
The on-shell t-matrix, $T_{\eta N}(E)$ is written
in terms of the scattering amplitude, $F_{\eta N}(E)$, 
using for the latter the effective range expansion,
\begin{equation}
F_{\eta N}(E)^{-1} + iq_\eta =\frac{1}{a} + 
\frac{r_0}{2}q_\eta ^2+sq_\eta ^4,
\end{equation}
where $q_\eta $ is the eta momentum corresponding to energy E. 
$t_{\eta N} (E)$ in Eq.~(\ref{greenequation} )
is written using another effective range expansion, namely, 
$f_{\eta N}(E) + i q_{\eta} v(q_{\eta})^2 = (1/a^s) + (r_0^s q_{\eta}^2/2) 
+ (s^s q_{\eta}^4)$ where the parameters $a^s$, $r_0^s$ and $s^s$ are related 
to $a$, $r_0$ and $s$.  
The scattering amplitude and the t-matrix are related as, 
$f_{\eta N}(E) =-\frac{\sqrt{s _{\eta N}}}{2\pi}t_{\eta N}(E)$.
The best set of the ``effective range expansion" 
parameters given in \cite{etaN2} is: 
$a$=0.91 + $i$ 0.27 fm; $r_0$ = -1.33 - $i$ 0.30 fm; $s$ = 
-0.15 - $i$ 0.04 fm$^3$.

In another class of phenomenological studies, again coupled-channel analyses of the four channels, $\gamma N$, $\pi N$, $\pi \pi N$,
and $\eta N$ are performed, but the T-matrices are now constructed microscopically 
in dynamical coupled-channel models of meson production reactions 
including nucleons and their resonances. 
The first comprehensive calculation in this class was done in \cite{bhalerao}. 
The authors considered the $\pi N$ collision channels, 
$\pi N\rightarrow \pi N$, $\pi N\rightarrow \pi \pi N$, and $\pi N\rightarrow \eta N$, in the energy region $\sqrt {s}$=1488 to 1600 MeV. 
They used a separable interaction model and assumed the reactions to 
proceed via the $N^*(1535)$ or $\Delta (1232)$ isobars. The interaction 
satisfied a Lippmann-Schwinger (LS) equation
\begin{equation}
T=V+VG_0T
\end{equation}
with the transition interaction $V_{ij}$ for channel $i\rightarrow j$ for a given partial wave $l$ and for a 
specific baryon resonance  doorway state $\alpha $   given by
\begin{equation}
\langle p\prime |V_{ij}^{\alpha l}(\sqrt {s})|p\rangle =\frac{h_{i\alpha }^l(p\prime)h_{\alpha j}^l(p)}
{\sqrt {s}-m_\alpha -\Sigma _{2\pi }^\alpha (\sqrt {s})},
\end{equation}
with $h$ being the vertex function. 
It had the strength $g$ of the coupling and the range $\Lambda $ of off-shell extrapolation as parameters. 
The full solution of the coupled channel LS equation was given as 
\begin{equation}
\langle p\prime |T_{ij}^{\alpha l}(\sqrt{s})|p\rangle =\frac{h_{i\alpha }^l(p\prime)h_{\alpha j}^l(p)}
{\sqrt {s}-m_\alpha -\Sigma _\pi ^\alpha (\sqrt {s})-\Sigma _\eta^\alpha (\sqrt {s})-\Sigma _{2\pi }^\alpha (\sqrt {s})},                   
\end{equation}    
where $\Sigma _x$ corresponds to the self energy of the particle $x$. 
The parameters of the model were 
$g$, $\Lambda $, and the bare mass $m_\alpha $. 
These parameters were determined by fitting the $\pi N$ phase shifts. Cross sections for the $\pi ^-p\rightarrow \eta n$ channel were 
predicted, which agreed very well with the then available data. The s-wave $\eta N$ scattering lengths obtained from the $\eta N$ scattering 
amplitude from these calculations were 
0.27 + $i$ 0.22 fm and 0.28 + $i$ 0.19 fm for two available sets of 
$\pi N$ data. The S$_{11}$ $\eta N$ 
phase shifts were found to have positive values, 
indicating an attractive interaction.

Another study in the above class, which  is quite  recent, is a detailed investigation of the $\pi ^-p\rightarrow \eta n$ in Ref. \cite {etaN4}. 
Unlike \cite {bhalerao} this approach fixes all the parameters of the model by  fitting all   
available data on the $\pi ^-p\rightarrow \eta n$ reaction  from threshold 
up to total centre-of-mass 
energy of about 2 GeV. Constraints on the model parameters 
from the $\pi N$ elastic scattering are incorporated from the earlier 
studies of this reaction by this group. The meson baryon (MB) channels 
included in the studies are $\pi N$, $\eta N$, $\pi \Delta $, $\rho N$, and $\sigma N$. The MB transition amplitudes in each partial wave 
is written as a sum of the background term and a resonance term. 
The former is taken energy independent and the latter is taken to have 
the usual resonance structure with the $N^*$ propagator, $D_{ij}(E)$,  with the associated resonance self-energy, sandwiched between 
the dressed vertex functions, $\Gamma _{MB\rightarrow N^*}$'s. The calculations include nine nucleon resonances, though the final results 
of the calculations show that the dominant contribution to the $\pi ^-p\rightarrow \eta n$ reaction comes from the $S_{11}(1535)$ resonance. 
 This paper presents the t-matrix for each channel 
in all partial waves. 
The value of the scattering length from the calculated s-wave $\eta N$ 
T-matrix comes out to be $a_{\eta N}$=0.30 + $i$ 0.18 fm. 
It is very interesting to note that this value 
agrees very closely  with the value obtained about 
two decades back in \cite {bhalerao} using a similar theoretical framework (presented in the previous paragraph). 

\subsection{Theoretical}
Theoretical studies have also been done in literature in the framework of chiral perturbation theory ($\chi $PT) 
\cite {etaN5, kww, lz, krip1}. $\chi $PT is an effective field theory which maintains the basic symmetries of QCD. It describes 
well the interaction between the pseudoscalar 
($J^{\pi}$ = 0$^-$) mesons and the ground state baryon octet in an almost parameter free way. 
The transition interaction, $V_{ij}$ between different meson-baryon 
channels in the lowest order chiral Lagrangian is given by
\begin{equation}
V_{ij}=-C_{ij}\frac{1}{4f_if_j}(2\sqrt {s}-M_i-M_j)\times \sqrt {\frac{M_i+E_i(\sqrt {s})}{2M_i}}\sqrt {\frac{M_j+E_j(\sqrt {s})}{2M_j}}
\end{equation}
where $E_i (E_j)$ is the energy of the incoming (outgoing) baryon 
and $M$ denotes the mass of the baryon. $f_x$ is the weak decay 
constant of the meson $x$. $C_{ij}$ is a fixed constant for 
each transition, reflecting the SU(3) symmetry. 
The scattering matrices for different channels are obtained by 
solving the coupled-channels Bethe-Salpeter 
equations. Various meson baryon 
channels (with $0^-$ mesons and $1/2^+$ octet baryons) 
which couple to strangeness S=0 are $\pi^-p$, $\pi ^0n$, $\eta n$, 
$K^+\Sigma ^-$, $K^0\Sigma ^0$ and $K^0\Lambda $. 
The values of the scattering length, $a_{\eta N}$ 
obtained in this approach in \cite {etaN5, kww, lz} 
are 0.20 + $i$ 0.26 fm, 0.26 + $i$ 0.25 fm and 0.18 + $i$ 0.42 fm, 
respectively. In \cite{krip1}, however, 
a much larger value as compared to these, namely, 
$a_{\eta N}$ = 0.54 + $i$0.49 fm was obtained. 

Finally, we conclude  that the best possible calculations using different models and including all available experimental data 
have been done to obtain the $\eta N$ scattering amplitude. Nevertheless, 
no definite value for the $\eta N$ scattering length is obtained. 
There exists a large spread \cite{etaN2, abm, sibir} in the values. 
This happens essentially 
because of the inherent nature of the problem that $a_{\eta N}$ can not 
be extracted directly from the experimental data. We also observe that, 
in general, the values of  $a_{\eta N}$ coming from theoretical models, like $\chi $PT and microscopic phenomenological models tend 
to be smaller than those obtained phenomenologically from the experimental data directly.
A table of all values of $a_{\eta N}$ obtained until 2002 can be found in 
\cite {sibir}. Apart from these and the ones mentioned in the discussions 
above, there exist calculations with large values such as 
0.991 + $i$ 0.347 fm \cite{penner}, 1.03 + $i$ 0.49 fm \cite{lutz} and 
a rather small $a_{\eta N}$ = 0.41 + $i$ 0.26 fm \cite{gaspa}. 

\section{Searches for unstable eta-mesic nuclei}
The existence of an eta-mesic nucleus, i.e., a quasibound state 
of the $\eta$ meson and a nucleus was predicted due to the attractive 
nature of the $\eta N$ interaction. Since its first mention in 1986, several 
experimental and theoretical searches have been performed for light as 
well as heavy eta-mesic nuclei. The experimental searches involve 
the production of $\eta$ mesons and hence signals for the existence 
of eta-mesic states via their possible decay modes and final state 
interactions of eta mesons 
with nuclei. The theoretical works concentrate on the 
calculation of the eta-nucleus elastic scattering amplitudes 
and the solutions with $\eta$-nucleus potentials using 
different approaches. Possible signals from experiments and the conclusions 
drawn from theoretical works will be discussed in the subsections below.

\subsection{Experimental searches}
\begin{figure}[h]
\centering
\includegraphics[width=8cm]{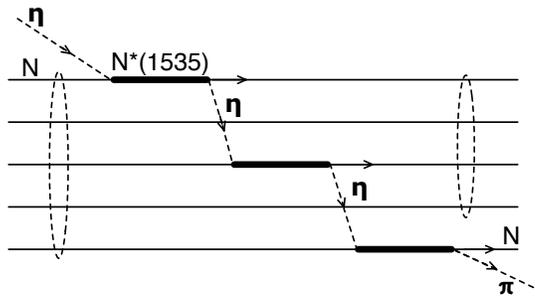}
\caption{Diagrammatical representation  of a series of excitations and 
decays of the $N^*$$(1535)$ resonance in the nucleus.}\label{nstar_inmedium}
\end{figure}

Due to the attractive $\eta N$ interaction and the proximity of its threshold to the mass of the $N^*(1535)$  resonance, 
the $\eta$-nucleus interaction can be considered as a series of  excitations 
and decays of $N^*(1535)$ on the different
constituent nucleons, which eventually 
decays to the $\pi N$ channel, as shown in Figure~\ref{nstar_inmedium}. The strategy of the experiments looking for
 $\eta$-mesic nuclei is to focus  on  the $\pi N$ pair coming from the $N^*(1535)$ resonance decaying in the nucleus.

The first experimental search of the $\eta$-nucleus bound states  \cite{firstexp} was performed at
 the Brookhaven National Laboratory (BNL) by studying pion collisions with lithium, carbon, oxygen 
 and aluminium but the results  turned out to be negative. 
A few years later another attempt to find 
 these exotic states was made at the Los Alamos Meson Physics 
Facility (LAMPF) \cite{lampf} by 
 investigating the process: $\pi^+$  $^{18}$O  $ \to$  $\pi^-$  $^{18}$Ne 
by varying the beam energy from 350-440 MeV, for the momentum transfer 
ranging from 0-210 MeV. This experimental search for an $\eta$-mesic state in 
a double charge exchange (DCE) 
reaction was motivated by an earlier work of Haider and Liu 
\cite{hailiudblchex} where the authors 
studied the $^{14}$C ($\pi^+, \pi^-$)$^{14}$O reaction 
theoretically and predicted a resonance structure in the excitation 
function of this reaction at a pion kinetic energy of 419 MeV. 
For pion beam energies of around 400 MeV, the DCE reaction can proceed 
in a nucleus 
via the $\pi^+ \to \pi^0 \to \pi^-$ process (where $\pi^+ n \to \pi^0 p$ is 
followed by $\pi^0 n \to \pi^- p$)
as well as the 
$\pi^+ \to \eta \to \pi^-$ process. The $\eta$ can either be in the 
continuum or in a strongly bound $\eta$-nuclear state. They showed that 
the DCE amplitude associated with the bound $\eta$ possessed a resonance 
structure with a narrow width ($\sim$ 10 MeV). Thus an $\eta$-nucleus 
state could act as a doorway state for the ($\pi^+,\pi^-$) reaction 
channel. The situation is somewhat analogous to the appearance of a
resonance structure due to compound nucleus formation 
in low energy nuclear reactions.

The poor statistics in the LAMPF experiment, however, did not
   allow to conclude more than a  weak affirmation of the presence of 
a structure near the eta threshold.
Since then several experiments have been made in different laboratories studying different kinds of eta 
producing reactions and some experiments claim to have found $\eta$-mesic 
nuclei. In this subsection we will review the different strategies followed in the previous 
  investigations and the results obtained in the same, as well as the new upcoming experiments and their assets. 

\subsubsection{Eta meson production in proton induced reactions} \label{expsrch1}
Eta production with proton beams scattered on different light targets ($p$, $d$, $^3$He)
 has been studied experimentally and the resulting data in all these processes show
  sharply rising amplitudes as the energy approaches the threshold region. This is an
   indication of the strong  attractive $\eta-A$ interaction and can be related to the 
    formation of a quasibound state in these systems. 

\begin{figure} [h]
\centering
\includegraphics[height=9cm,width=8cm]{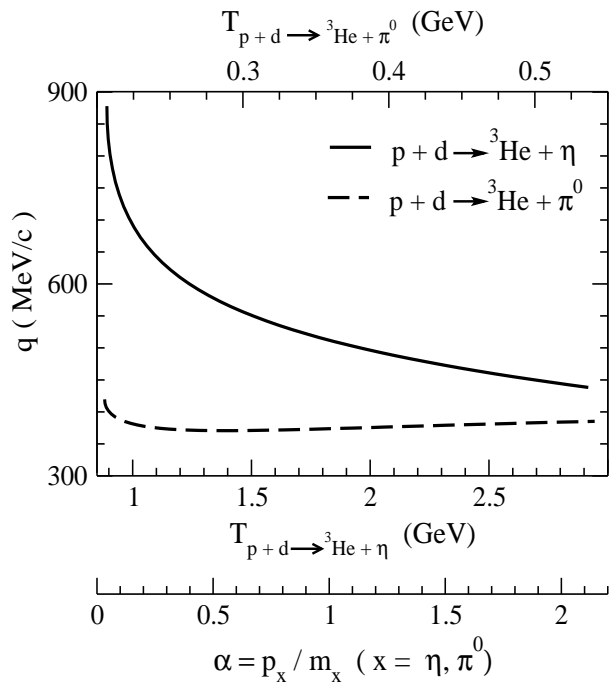}
\caption{A comparison of the momentum transfer involved in the
$p$  $d$ $\rightarrow$  $^3$He $\eta$ (solid line) and
the $p$  $d$ $\rightarrow$  $^3$He $\pi^0$ (dashed line and upper scale)
reaction
as a function of corresponding beam energies. The variable, $\alpha$, has been defined
such that there exists a common scale for the two reactions.} \label{momtrans}
\end{figure}

The reaction which has been studied most extensively  is $p$ $d$ $\to$ $^3$He $\eta$. 
The oldest data set on this reaction is available from the Laboratoire National Saturne, Saclay \cite{berth}. 
This study involved a 
measurement of the $p$  $d$ $\rightarrow$  $^3$He $\eta$ reaction
corresponding to $\eta$-production at very backward angles and at beam energies
ranging from 0.92 GeV to 2.6 GeV. Interestingly, the measured cross sections turned out to be
comparable to that of the $p$  $d$ $\rightarrow$  $^3$He $\pi^0$ reaction.
Such large cross sections were not expected since the 
$p$  $d$ $\rightarrow$  $^3$He $\eta$ reaction  
involves much larger momentum transfer (see Figure~\ref{momtrans} for a comparison of the momentum
transfer as a function of beam energy for these two reactions).
It can be seen from Figure~\ref{momtrans} that the momentum transfer for the 
$p$  $d$ $\rightarrow$  $^3$He $\eta$ reaction (solid line) is about 900 MeV
and for the
$p$  $d$ $\rightarrow$  $^3$He $\pi^0$ reaction (dashed line) 
is about 400 MeV at the respective thresholds.

\begin{figure} [h!]
\centering
\includegraphics[width=10cm]{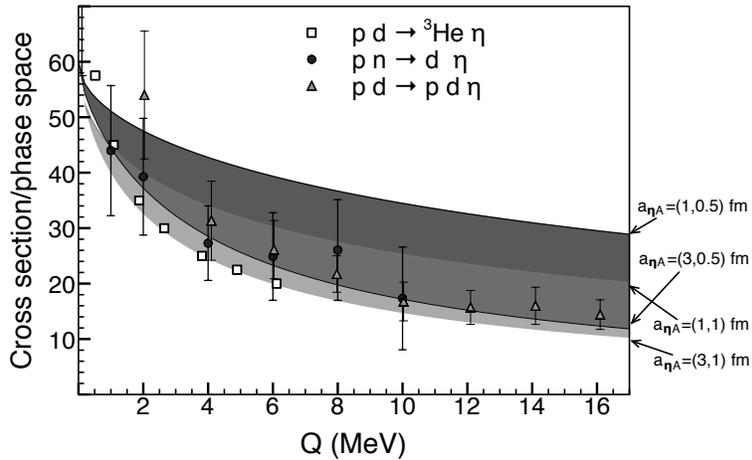}
\caption{Squared amplitude data on the $p$  $d$ $\rightarrow$  $^3$He $\eta$ (empty squares) \cite{dataMayer}, 
$p$  $n$ $\rightarrow$  $d$ $\eta$ \cite{calen} (filled circles) and
$p$  $d$ $\rightarrow$  $p$ $d$ $\eta$  (open triangles) \cite{hib00} reactions as a function of the excitation energy 
in the $\eta$-nucleus system. These amplitudes have been 
multiplied by arbitrary factors to facilitate a comparison among them. As can be seen, the slope of the  three data sets
 is very similar. The meaning of the the dark and light shaded region is explained in the text.
} \label{fsq}
\end{figure}

\begin{figure} [h!]
\centering
\includegraphics[width=9cm]{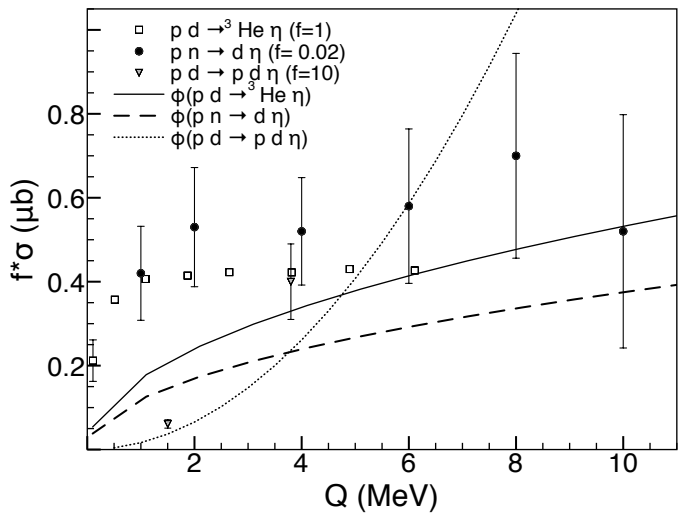}
\caption{A comparison of the total cross sections, multiplied by a factor f which is given in the figure legends, of the
$p$  $d$ $\rightarrow$  $^3$He $\eta$ \cite{dataMayer}, $p$  $n$ $\rightarrow$  $d$ $\eta$ \cite{calen} and
$p$  $d$ $\rightarrow$  $p$ $d$ $\eta$  \cite{hib00} 
reactions as a function of the excitation energy 
($Q = \sqrt{s} - \sum_{i} M_f^i$, 
where $s$ and $M_f^i$ are the total energy of the system and the mass of the $i$th particle in the final state). The lines indicate the phase space, $\phi$, 
for the three reactions.} \label{totxn}
\end{figure}

Another set of measurements on the same reaction  \cite{dataMayer} was later made at the same laboratory
which focused on energies very close to threshold. 
The $p$  $d$ $\rightarrow$  $^3$He $\eta$ cross sections
were measured at eight different proton energies from 0.2 to 11 MeV above threshold. Apart
from being in agreement with the observation of the previous experiment 
regarding the surprisingly large cross sections, the
experimental data in this region showed that the squared-amplitude decreased nearly by a factor of four within this small 
energy region, which corresponds to an increase
 in the $\eta-^3$He center of mass momentum by about 75 MeV/c. 
The forward-backward asymmetry was negligible indicating that the 
angular distributions were isotropic in the center of mass system. In other words,
the final state particles were found to be produced in the s-wave. Similar features were
also found in the data for the $p$  $n$ $\rightarrow$  $d$ $\eta$ \cite{calen} and $p$  $d$ $\rightarrow$  $p$ $d$ $\eta$ \cite{hib00}
reactions. The squared amplitudes and the total cross section data for these three reactions, in the threshold region, are shown in 
Figs.~\ref{fsq} and \ref{totxn} as a function of the excitation energy $Q = E_{\eta A} - M_\eta - M_A$, where $E_{\eta A}$, $M_\eta$, $M_A$
refer to the total energy of the $\eta$-nucleus system, the mass of the $\eta$-meson and the mass of the nucleus, respectively. Different data sets 
in Figure~\ref{fsq} have been multiplied by arbitrary factors for the sake of comparison and it can be
seen from this figure that the slope of the different squared amplitudes of $\eta$-light nuclei is very similar. The sharp structure
of the amplitudes near threshold is a clear
 manifestation of the strong $\eta$-nucleus final state interaction  which hints towards the existence of
 possible quasi bound states in such systems. 
Assuming the dominance of the $\eta-A$ interaction in these processes, 
the corresponding squared amplitude data is often fitted 
using the $\eta-A$ scattering length, $a_{\eta A}$, as \cite{watsonbook}
\begin{equation}\label{slef}
\mid F (k) \mid^2 =  \frac{f_B}{ \mid 1 - ik a_{\eta A}\mid^2},
\label{scatlen_fit}
\end{equation} 
where $k$ is the momentum in the $\eta-A$ center of mass system and 
$f_B$ is a normalization factor which is related to 
the contribution of the Born amplitude of the reaction. 
The fitting procedure usually concentrates on reproducing the shape of the 
data with $f_B$ being an arbitrary parameter used to reproduce the right 
magnitude of the data. 
The fitted value of the $\eta-A$ scattering length
 is then used to infer an indirect evidence of the formation of 
an $\eta-A$ quasibound state. We show the result of the calculation 
of Eq.~(\ref{scatlen_fit})
by taking  $1\leq |\Re e \{a_{\eta A}\}| \leq 3$ for $\Im m\{a_{\eta A}\} = $ 0.5(1) by dark(light) shaded regions 
(these values lie within the range of
$a_{\eta A}$ extracted in Ref.~\cite{sibirtsev}). It can be seen that a
 reasonable fit is obtained for  larger values of $a_{\eta A}$. Notice that Eq.~(\ref{scatlen_fit}) is blind to the sign of the
  $\Re e\{a_{\eta A}\}$. Though 
the above approximation looks promising and is often used in literature, 
such a fitted value of 
the eta-nucleus scattering length can be quite misleading. 
We shall discuss this point in greater detail in Chapter 5. 

Furthermore, an indication of the presence of a 
quasibound/quasiresonant state can be also seen from  the total cross section data (shown in Fig~\ref{totxn}) 
 which cannot be explained by the phase space of the reactions (shown as solid, dashed and dotted lines for $p$~$d$ $\rightarrow$~
 $^3$He~$\eta$, $p$~$n$~$\rightarrow$~$d$~$\eta$ and $p$~$d$~$\rightarrow$~$p$~$d$~$\eta$ reactions, respectively).
 
Measurement of the cross sections for the proton induced $\eta$ production on light targets has been made with more statistics near the 
threshold region \cite{dataMersmann,dataRausmann,pd3He2,pd3He3,piskor}.
More data are also being made available at higher energies for these reactions \cite{pd3He4,parascatt}. The data from these experiments
 show anisotropic angular distributions indicating a possible interplay of 
higher partial waves in the reaction mechanism at energies farther from the 
threshold region.

More recently, differential cross section measurements for 
$^6$Li($p$,$\eta$)$^7$Be reaction at 11.28 MeV excess energy have been performed  
at COSY \cite{ulc,gemcosy} where the missing mass of the recoiling 
nucleus $^7$Be was measured with the aim of studying the FSI. 
The total cross section is found to be around 
(8.6 $\pm$ 2.6 stat. $\pm$ 2.4 syst.) nb. 
Compared to this the total cross sections around the same excess energy 
for $p d \to\, ^3$He $\eta$    and $d d \to \alpha\, \eta$ 
are  around (407 $\pm$ 20) nb \cite{dataMersmann}
and (16 $\pm$ 1.6) nb \cite{dd1,dd2}, 
respectively, showing thereby a drastic decrease in the $\eta$ 
production cross section with an increase in the target nucleus mass number.

A search for the binding of the eta mesons in heavier nuclei has been 
done recently at COSY in the $^{27}$ Al ($p$, $^3$He) $p$ $\pi^-$ $X$ 
transfer reaction \cite{cosy}. 
The $^3$He is detected at 0 deg. and the beam energy is 
chosen such that the eta meson goes into the nucleus 
($^{25}$ Mg  in this case) with nearly no momentum, 
increasing thereby its binding probability (recoil free kinematics). 
Considering that the bound eta meson in the nucleus decays through the 
$\pi p$ channel, a triple coincidence is performed in the experiment, 
where $p \pi^-$ going in a cone around back to back are measured 
in coincidence with  the $^3$He nucleus having missing mass 
around 550 MeV. This study has led to the finding of a peak structure 
around 13 MeV with a half width around 5 MeV. This has been attributed to 
an eta-nucleus bound state in $^{25}$Mg. 
An upper limit on the total cross section for this channel has been 
fixed in this experiment around 0.5 nb.

A theoretical analysis of the above data has been done very recently, 
which strengthens this conclusion (discussed further in Section 3.2.2.).
For an overview of the experiments planned at COSY, see Ref. \cite{moskal}. 

\subsubsection{Photoproduction and quasibound states }\label{expsrch2}
Almost a decade after  the studies reported in Refs~\cite{firstexp,lampf},  very conclusive results regarding a clear evidence  of 
an $\eta$-mesic  nucleus were released from the Mainz Microtron facility (MAMI) \cite{taps}, where the $\eta$-meson was claimed
 to form a quasibound resonance with a nucleus as light as $^3$He.
The inclusive cross section for the $\gamma$~$^3$He~$\to$~$\eta$~$X$ reaction was measured with the beam energy ranging 
from threshold to 820 MeV and a resonant structure was found above the $\eta$ production threshold in the coherent 
$\gamma$ $^3$He $\to$  $\eta$ $^3$He  cross sections while a peak was seen slightly below the threshold in the $\pi^0 p$ decay 
channel. The mass and width of the resonance extracted from these cross sections were reported to be
 $[(-4.4 \pm 4.2) - i( 25.6 \pm 6.1)]$ MeV.  These results are in striking coincidence with those reported in Ref.~\cite{ourtimedelay} 
 for small $\eta N$ scattering lengths, where a possibility of the existence of the $\eta-^3$He quasibound  state was studied by 
 calculating the Wigner's time delay method using the amplitudes obtained by solving the few body equations for the system. 
 However, this work suffered from the presence of a pole 
near the threshold which refrained the authors from making strong claims. 
 This work was revisited and a more refined investigation was later 
made \cite{meprl} by subtracting the singularity close to 
  threshold. This led to the finding of the 
existence of $\eta \,^3$He quasibound to quasivirtual states 
when the $\eta N$ scattering lengths were varied 
from small to large values 
($a_{\eta N}$ = (0.28, 0.19) fm and (0.88, 0.41) fm which correspond 
to $a_{\eta^3{\textrm He}}$ = (1.16, 0.88) fm and 
(2.14, 5.71) fm respectively). A more detailed discussion of these
calculations will be presented in the next subsection.

 However, soon after the release of Ref.~\cite{taps} a critical comment on the same was published \cite{hanhartcomment}
  which analysed the $\gamma$~$^3$He~$\to$~$\pi^0$~$p$~$X$ data of Ref.~\cite{taps} and it was  shown that the data 
  was more compatible with the formation of a virtual $\eta-^3$He state rather than with a bound state if the corresponding 
  scattering length was assumed to be $ a_{\eta ^3{\textrm He}} = (\pm 4, 1 )$ fm or (0, 3.5) fm. The comment was followed by a 
  reply from the experimental group at MAMI \cite{tapsreply} where a more refined binning of the data was done. This led to results
   similar to Ref.~\cite{taps} and it was concluded that a better statistics would lead to a more unambiguous interpretation.  
   To resolve the issue, a measurement of the cross sections for 
the coherent  $\gamma$ $^3$He $\to$  $\eta$ $^3$He reaction was 
repeated \cite{mami2012} with better statistics and by measuring 
the $\eta \to 2\gamma$ as well as  $\eta \to 3\pi^0 \to 2\gamma$ decay 
channels, in order to have better control on 
the systematic uncertainties. 
The latter channel 
was not measured in Ref.~\cite{taps}. 
   Unfortunately, this more refined study did not lead  to more conclusive results. In fact the peak found in Ref.~\cite{taps} was reproduced 
   in this new experiment but it was found that the peak cannot be attributed 
to the existence of a resonance unambiguously 
since it can be explained as an artifact arising from the quasifree 
pion production background \cite{fujioka}.

 The photoproduction of $\eta$-mesic nuclei has also been studied on  $^{12}$C at the Lebedev Physical Institute \cite{rusos}
  and a lowering of the mass of the $S_{11}$ resonance has been found in the $\pi^+ n$ spectrum when the relative angle between 
  the two particles is 180$^0$.  This has been interpreted as the indication of the formation of an $\eta$-mesic nucleus. Photoproduction
   of the $\eta$-meson is also being studied on lighter nuclei at other laboratories \cite{cbelsa}.

Finally, it should be mentioned that  production of the $\eta$-mesons is also planned  at the DA$\Phi$NE facility through the radiative 
decays of the $\phi$ mesons \cite{frascati}, with one of the motivations being  finding the $\eta$-mesic nuclei.

\subsubsection  {Eta production with pion beams}\label{expsrch3}
In spite of the failure of the earlier experimental searches of $\eta$-mesic systems with pion beams at the BNL and LAMPF
 facilities \cite{firstexp,lampf}, a proposal to study the pion induced $\eta$ production  on the $^7$Li and $^{12}$C targets 
 has been made recently at J-PARC \cite{fujioka}.  However, in contrast with the previous investigations \cite{firstexp,lampf}, there are plans to carry out 
these experiments within the recoilless kinematics and by demanding an exclusive measurement
   (like the studies at MAMI and COSY \cite{taps,cosy}). The beam momentum for these experiments is planned to vary in the range
    0.7-1 GeV in order to produce recoilless $\eta$-mesons \cite{hideko}. 
Furthermore, in order to deduce the possible background from the
     quasifree production (without excitation of the $N^*(1535)$ resonance in the medium) of $\pi^-$~$p$ pair present in the 
     corresponding data, it is planned to measure the cross sections for 
the $\pi^+$~$d$~$\to$~$p$~$p$~$\eta$ reaction also. Theoretical analyses 
of the pion induced eta production within the distorted wave impulse 
approximation can be found in \cite{dwia}.

Before ending  the discussions on the experimental investigation of $\eta$-mesic nuclei, it should be mentioned that $\eta$ production
 has also been measured in the  $d$~$d$~$\to$~$\alpha$~$\eta$ \cite{dd1,dd2} reaction. However, the finding of a $\eta-\alpha$
  quasibound state was not confirmed in these studies.

This discussion can be summarized by stating that some of the existing experimental studies  do claim 
 the existence of $\eta$-mesic nuclei. However, none of these results have been  confirmed by other independent experiments. 
 Many new experiments are planned which promise better statistics and less ambiguity from the background, which in turn will 
 help in making more definitive conclusions regarding the existence of $\eta$-mesic nuclei in future.

\subsection{Theoretical studies}
The theoretical approaches used in the search for the existence of 
quasibound states of eta mesons and nuclei can be broadly classified into 
two categories: ones using models based on few body equations 
for systems of eta mesons and 2 to 4 nucleons and others 
based on optical potentials for the search of heavy eta-mesic nuclei.
In what follows, we shall discuss the findings 
of the works representative of each category separately. The existence 
of a quasibound state in these works 
is confirmed by the occurrence of an S-matrix  
pole in the complex momentum plane, 
Argand plots of scattering amplitudes or prominent peaks in time delay. 
The interesting features of the time delay method which can be used as 
a complementary tool for locating resonances, quasivirtual and quasibound 
states will also be discussed below. In Figure 5 we show the location of 
the poles of the various states in the complex planes to clarify the 
notation used in this report. The physical and 
unphysical sheets correspond to $\Im$m p $>$ 0 and $<$ 0 respectively. 
\begin{figure}[h]
\centering
\includegraphics[width=8cm,height=6cm]{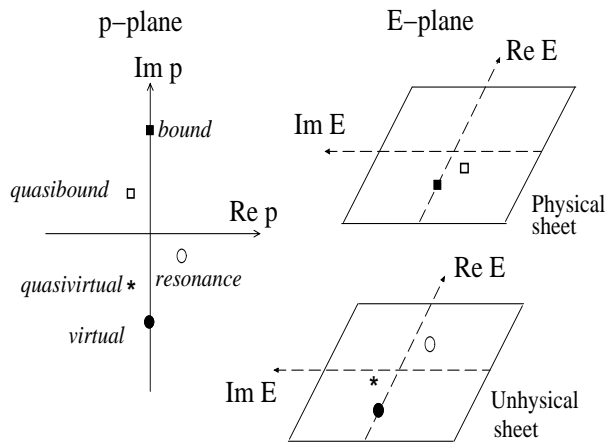}
\caption{\label{fig:eps1} Positions of the poles in the complex energy 
and momentum planes corresponding to bound, virtual, resonant, quasivirtual 
and quasibound states.}
\end{figure}

\subsubsection{Few body equations and unstable states of light nuclei}
One of the first predictions for light eta-mesic nuclei using few body 
equations was made by T. Ueda \cite{ueda}. Since the $S_{11}$ resonance decays 
prominently to $\pi N$ or $\eta N$, the author assumes that the $\eta N N$ 
system couples necessarily to the $\pi N N$ system. The three body equation 
for the amplitude $X_{\alpha \beta}$ is written as
\begin{equation}
X_{\alpha \beta} = Z_{\alpha \beta} + \sum_{\mu,\nu} Z_{\alpha \mu}\, 
\tau_{\mu \nu}\, X_{\nu \beta}
\end{equation}
where $Z$ and $\tau$ are given by form factors of separable input potentials 
and the free three body Green's functions \cite{uedaPLB}. $Z_{\alpha \beta}$ 
is the particle rearrangement term between particle channels $\alpha$ and 
$\beta$ and $\tau_{\mu \nu}$ is the propagation term with a spectator 
particle and an interacting pair in the intermediate channel. Concentrating 
on isospin 0 and $J^{\pi} = 1^-$, the author considers the coupled 
$\eta N N$ - $\pi N N$ system and evaluates the $\eta d$ scattering amplitude.
The input two-body interactions involve the $\pi N$ potential in the $P_{11}$ 
state (though the Roper resonance is ignored), $N N$ in $^3S_1$-$^3D_1$ state 
and the $\pi N \to \eta N$ potential in the $S_{11}$ state. A pole structure 
corresponding to a quasibound state 
is observed in the amplitude at $\Re$e E$_{\eta}$ = 
- 2 MeV and $\Im$m E$_{\eta}$ = -10 MeV.

Following this prediction, a few years later Rakityansky et al. \cite{rakit2}
investigated the possibility of forming $\eta d$, $\eta$-$^3$H, 
$\eta$-$^3$He and $\eta$-$^4$He states using few body equations.  
Within a finite rank approximation (FRA), i.e.,retaining the interacting 
nucleus in its ground state, they studied the movement of the poles of the 
elastic $\eta$- light nucleus scattering amplitudes. The $\eta$-nucleus 
transition matrix at any complex energy $z$ was evaluated by solving the 
integral equation
\begin{eqnarray}
T(\vec{k}^{\prime}, \vec{k}, z) &=& \langle\vec{k}^{\prime}, \psi_0| 
\sum_{i=1}^{A} T^0_i(z)|\vec{k}, \psi_0 \rangle \\ \nonumber
&+&\, \epsilon_0\int 
{d^3k^{\prime\prime} \over (2\pi)^3} {\langle\vec{k}^{\prime}, \psi_0| 
\sum_{i=1}^{A} T^0_i(z)|\vec{k}^{\prime\prime}, \psi_0 \rangle \over 
\biggl ( z - {k^{\prime\prime 2} \over 2\mu} \biggr ) \biggl( z - \epsilon_0 
- {k^{\prime\prime 2} \over 2\mu} \biggr)} \, T(\vec{k}^{\prime\prime}, 
\vec{k}, z)\,
\end{eqnarray}
where 
\begin{equation}
\langle\vec{k}^{\prime}, \psi_0| \sum_{i=1}^{A} T^0_i(z)|\vec{k}, \psi_0 
\rangle = \int d^{3(A-1)}r |\psi_0(\vec{r})|^2 \sum_{i=1}^A T_i^0
(\vec{k}^{\prime}, 
\vec{k}; \vec{r}; z)\, .
\end{equation}
$T_i^0(\vec{k}^{\prime}, \vec{k}; \vec{r}; z)$ is obtained by 
solving another integral equation, namely, 
\begin{eqnarray}
T_i^0(\vec{k}^{\prime},\vec{k};\vec{r},z) &=& t_i(\vec{k}^{\prime},\vec{k};
\vec{r},z) \\ \nonumber
&+& \int {d^3k^{\prime\prime} \over (2\pi)^3} {t_i(\vec{k}^{\prime},
\vec{k}^{\prime \prime};\vec{r},z) \over z - {k^{\prime \prime}\over 2 \mu}}
\, \sum_{j \ne i} T_j^0(\vec{k}^{\prime \prime},\vec{k};\vec{r},z)
\end{eqnarray}
where $t_i(\vec{k}^{\prime},\vec{k};\vec{r},z) = t_{\eta N} (\vec{k}^{\prime}, 
\vec{k},z) \, exp[i(\vec{k} - \vec{k}^{\prime}) \cdot \vec{r}]$ with 
$t_{\eta N} (\vec{k}^{\prime}, \vec{k},z)$ being the off-shell $\eta N$ 
amplitude. This amplitude is written using a separable form and assuming the 
dominance of the $S_{11}(1535)$ resonance. The input $\eta N$ scattering 
length, $a_{\eta N}$ was varied and complex poles 
corresponding to quasibound states of 
$\eta d$, $\eta$-$^3$H, $\eta$-$^3$He and $\eta$-$^4$He were found for 
$\Re e a_{\eta N}$ in the range [0.27, 0.98] fm. The above formalism was later 
used to study the effects of the final state interactions in $\eta$ producing 
reactions such as the $p d \to$ $p d \eta$, $p d \to \, ^3$He $\eta$ and 
$p \,^6$Li $\to \, ^7$Be $\eta$ 
\cite{kanchan1,kanchan2,kanchan3,neelam1,neelam2}. 

A more rigorous treatment of the $\eta d$ amplitude using the 
Alt-Grassberger-Sandhas (AGS) equations was carried out in \cite{shev1} using 
various values of the input $\eta N$ scattering length. The exact AGS 
results were compared with the approximate calculations involving FRA and 
multiple scattering theory (MST) \cite{greenmst}. It was found that both 
MST and FRA fail to give the $\eta d$ scattering length as obtained 
from the exact AGS calculations in the 
case of a strong $\eta N$ interaction (i.e., $\Re e\, a_{\eta N} >$ 
0.5 fm) while for small values of $\Re e \, a_{\eta N}$ 
these methods work reasonably 
well. A three body resonant state near the $\eta d$ threshold 
was found in \cite{shev2} by the same authors using AGS equations. The 
resonance moved toward the $\eta d$ threshold when $\Re e \,a_{\eta N}$ was 
increased and turned into a quasibound state at $\Re e \,a_{\eta N} $  
$\simeq$ 0.7 
- 0.8 fm depending on the choice of $\Im m \,a_{\eta N}$. 

Running counter to all the earlier claims of $\eta d$ quasibound states, 
a relativistic three body Faddeev 
calculation performed in \cite{garci1, garci2} did 
not find any such state. The authors claimed the existence of quasivirtual 
states only which moved farther away from threshold with decreasing values 
of $\Re e \,a_{\eta N}$. The authors attribute the difference in their 
conclusions as compared to earlier works to be due to the difference in 
the treatment of the two-body interactions which enter as an input to 
the three-body calculations. Details of the differences can be found 
in \cite{garci2,garci3}. In \cite{garci1} a parametrization of the 
$\eta d$ amplitude using the effective range formula was also provided. 
Using this parametrization, it was later shown in \cite{ourtimedelay} that 
the author in \cite{garci1} had indeed missed a pole at -17 MeV corresponding 
to a quasibound state of the $\eta d$ system using model $0$ in that
work. Thus the calculation in 
\cite{garci1} did support a quasibound $\eta d$ state though quite away from 
threshold. We conclude this discussion by mentioning yet another Faddeev 
approach \cite{deloff} which was used to evaluate the $\eta d$ and $K^- d$ 
scattering lengths and ruled out the existence of $\eta d$ quasibound 
states. 
\subsubsection{Optical potential approaches for heavy $\eta$-mesic nuclei}
In this category two classes of approaches exist, 
semimicroscopic and fully microscopic.
The first class includes using the ``$t\rho $" approximation to the 
$\eta $-nucleus  optical potential, and the
second class includes the QCD based unitarized $\chi PT$ and the 
quark-meson-coupling (QMC) construction of the
$\eta -$nucleus potential. Both the approaches then use these potentials 
in the Schr\"odinger or Klein-Gordon equation and search for the bound states.

Let us start with the optical potential approach of Haider and Liu 
which was introduced in 1986 \cite{firstheor1,firstheor2} and led to the 
prediction of $\eta$-mesic states with mass number A $>$ 10. In a later 
work \cite{opticalhaider} the authors used a similar formalism to perform 
an exploratory study of the effects of the off-shell $\eta N$ interaction and 
limitations of approximations using on-shell $\eta N$ scattering lengths. 
Here we briefly describe the theoretical framework used in \cite{opticalhaider} 
and go on to discuss some interesting consequences drawn in this work. 
The complex energy eigenvalue -$|\epsilon|$-i$|\Gamma|$/2 of an eta-nucleus 
quasibound state in this work is calculated by solving the momentum space 
three dimensional integral equation 
\begin{equation}\label{haiderliupot}
{\vec{k}^{\prime 2} \over 2 \mu} \psi(\vec{k}^{\prime}) \, +\, 
\int\, d\vec{k} \, \langle \vec{k}^{\prime} | V| \vec{k} \rangle 
\psi(\vec{k})\, =\, E \psi(\vec{k}^{\prime})\, .
\end{equation}
Here $\langle \vec{k}^{\prime} | V | \vec{k} \rangle$ are momentum space 
matrix elements of the $\eta$-nucleus optical potential $V$ with $\vec{k}$ 
and $\vec{k}^{\prime}$ denoting the initial and final $\eta$-nucleus 
relative momenta respectively. Eq.~(\ref{haiderliupot}) is covariant and leads 
to the advantage that $V$ can be related to the elementary $\eta N$ process 
by unambiguous kinematical transformations (see Ref. [29] in 
\cite{opticalhaider}). The first order microscopic $\eta$-nucleus optical 
potential has the form 
\begin{eqnarray}
\langle \vec{k}^{\prime} | V | \vec{k} \rangle &=& \sum_j \int d\vec{Q} 
\langle \vec{k}^{\prime}, -\vec{k}^{\prime}-\vec{Q} | 
t(\sqrt{s_j})_{\eta N \to \eta N}|\vec{k},-\vec{k}-\vec{Q}\rangle \\ \nonumber 
& &\times \, \phi_j^*(-\vec{k}^{\prime}-\vec{Q}) \phi_j(-\vec{k}-\vec{Q})\, ,
\end{eqnarray}
where $\phi_j$ is the nuclear wave function with the nucleon $j$ having 
momenta -($\vec{k}+\vec{Q}$) and -($\vec{k}^{\prime}+\vec{Q}$) before and 
after the collision with $\vec{Q}$ being the Fermi momentum. 
$\sqrt{s_j}$ is the total energy in the centre of mass 
frame of the $\eta$ and the nucleon $j$ and is given by 
$$s_j = \biggl [ m_{\eta} + m_N - |\epsilon_j| - {\vec{Q}^2\over 2 M_{c,j} } 
\biggl({m_{\eta} +m_A\over m_{\eta} + m_N} \biggr ) \biggr ]^2$$
where $|\epsilon_j|$ is the binding energy of the $j^{th}$ nucleon and 
$M_{c,j}$ the mass of the core nucleus resulting after the removal of 
nucleon $j$. The integration over $\vec{Q}$ thus requires the knowledge of the 
amplitude $t_{\eta N \to \eta N}$ at subthreshold energies. The authors 
present the binding energies and widths of the quasibound states of 
$\eta$ mesons and $^{12}$C, $^{16}$O, $^{26}$Mg, $^{40}$Ca, $^{90}$Zr and 
$^{208}$Pb within the above off-shell calculation. 

They also present another calculation within a factorization approximation 
(FA) where $t_{\eta N \to \eta N}$ is taken out of the integral in 
(\ref{haiderliupot}) and evaluated at an ad hoc momentum $<\vec{Q}>$. The 
$\eta N$ centre of mass energy $\sqrt{s}$ is assumed to be 
$\sqrt{s} = m_{\eta} + m_N - \Delta$ with $\Delta$ being an energy shift 
parameter. Performing calculations for $\Delta =0, 10, 20, 30$ MeV, they 
notice that for $\Delta = 30$ MeV the FA results come quite close to the 
full off-shell calculations. Interestingly, the downward shift parameter 
$\Delta$ that fitted the $\pi N$ scattering data was also found to be 
around 30 MeV. The downward shift implies that the $\eta N$ interaction 
in $\eta$-bound state formation takes place at energies about 30 MeV 
below the free space threshold. Such a shift can lead to a reduction in the 
$\eta N$ attraction inside the nucleus and hence models using the $\eta N$ 
interaction in free space could actually be overestimating the $\eta$-nucleus 
binding energy. 

The optical potential within the FA was recently used \cite{haider2010} 
to explain the missing mass spectrum obtained in the recoil free transfer 
reaction $p$($^{27}$Al, $^3$He) $\pi p^{\prime} X$ performed by the COSY-GEM 
collaboration. The kinematics in this experiment were chosen in order to 
search for the $\eta$-mesic nucleus $^{25}$Mg$_{\eta}$. The authors 
in \cite{haider2010} showed that 
the observed peak structure occurs due to coherent contributions from 
processes where an $\eta$ binds to $^{25}$Mg to form an intermediate 
$^{25}$Mg$_{\eta}$ or it emerges as a pion through 
$\eta p$  $\to$ $\pi^0 p$ scattering in $^{25}$Mg without forming a 
quasibound state. This quantum interference, the authors observe, gives a 
weaker binding (-8 to -10) MeV as compared to the experimental value of 
(-13.13 $\pm$ 1.64) quoted in \cite{cosy}.

Binding energies and widths of quasibound $\eta$-mesic nuclei $^{12}$C, 
$^{40}$Ca and $^{208}$Pb were calculated in \cite{opticaloset1} by 
evaluating the $\eta$ self energy which is related to the optical 
potential. Assuming that the $\eta N$ interaction is dominated by N$^*$(1535) 
the $\eta$ self energy was written as
\begin{equation}\label{osetpot}
\Pi(k) = {g_{\eta}^2\, \rho \over \sqrt{s} - M_{N^*}+ i (\Gamma(s)/2) 
- i \Im m\,\Sigma_{N^*}(k^0,\vec{k}) + Re\, \Sigma_N - Re\, \Sigma_{N^*}}
\end{equation}
where $g_{\eta}$ is the $\eta$NN$^*$ coupling constant, 
$\rho$ is the nuclear density, $M_{N^*}$ the mass of N$^*$(1535), 
$\Gamma(s)$ its free width and $\Sigma_{N^*}$ the N$^*$ self energy in the 
nuclear medium. The optical potential in a finite nucleus was obtained 
using the local density approximation (LDA). This potential generated 
quasibound states with very large widths. The authors concluded that it is 
unlikely that any narrow peaks corresponding to bound eta states in nuclei 
would be detected experimentally. Another calculation in a similar spirit 
was done in \cite{opticaloset2} where the $\eta$ self energy was calculated 
in a chiral unitary approach \cite{osetinoue}. The quasibound states were 
once again found to be with the half widths larger than the separation 
of the levels.

Quark-meson coupling (QMC) model is a mean field description of the nucleus
like Quantum  Hadrodynamics (QHD), except that the
quark substructure of hadrons  is explicitly implemented in it.
It uses  mean-field equations with
meson fields explicitly coupling with the quarks in the
hadrons, e.g. $q\bar{q}$ in the bag for the eta meson. 
These equations are solved self consistently to determine the in-medium
quark masses, interacting fields, and eventually the eta meson mass in 
the nucleus. If we denote this in-medium $\eta $ meson mass by
$m_\eta ^*(r)$, the eta-nucleus potential is given 
by $V_\eta (r)=m_\eta ^*(r)-m_\eta (r)$, where the unstarred mass is the 
eta mass in free space. This potential includes 
the effect of $\eta - \eta ^\prime $ mixing. 
However, the QMC model does not include the
imaginary part of the potential consistently. It is introduced from 
outside assuming a specific form, with its strength as a
free parameter.
Using this potential, single particle energies for the $\eta$ 
are obtained solving the Klein-Gordon equation.
The results are  given for several closed shell nuclei and also for 
$^6$He, $^{11}$B and $^{26}$Mg. This work concludes
that one should expect to find bound states in all these nuclei. 

Finally in passing, we mention an interesting work \cite{Jidoetal} where the 
N$^*$(1535) being the lowest lying baryon with parity opposite to that of
the nucleon is viewed as the chiral partner of the latter. 
It was found that the N$^*$-N mass gap decreases 
in the nuclear medium with increase in density (chiral symmetry restoration) 
and the calculations in \cite{Jidoetal} show the existence of two bound 
eta-nucleus states at about -80 MeV. Other calculations by the same authors 
based on chiral models can be found in \cite{otherJidos}. 

\subsubsection{Collision times of eta mesons and light nuclei}
In \cite{ourtimedelay,meprl} the search for quasibound states of light 
$\eta$-mesic nuclei was carried out using an approach based on Wigner's 
time delay \cite{wigner,eisen}. This method has been applied earlier to 
characterize hadron resonances \cite{hadrontimedelay1,hadrontimedelay2} at 
positive energies but was not used to locate quasibound and quasivirtual 
states. In \cite{ourtimedelay} the usefulness of the method is demonstrated 
through a pedagogical example of the neutron-proton system where one 
can also locate the bound and virtual states via time delay plots. 
The method is then applied to reproduce some quasivirtual $\eta$-mesic 
states already found in literature and predict quasibound states of 
$\eta d$, $\eta$-$^3$He and $\eta$-$^4$He nuclei. Here we briefly describe 
the concept of Wigner's time delay and a modification of it which is useful 
for locating $s$-wave states. 
A rather different approach based 
on the concept of
``dwell time" was introduced by F. T. Smith in 1960 \cite{smith}. 

Wigner showed the phase time delay in single channel elastic scattering 
$\tau_{\phi}(E)$ (where $E = \hbar^2 k^2/2 \mu$ is the particle energy and 
$\hbar k$ the momentum) 
to be 
related to the energy derivative of the scattering phase shift $\delta_l(E)$ 
by 
$\tau_{\phi}^l (E) = 2 \hbar \,{d\delta_l/dE}$.
The time delay in elastic 
scattering was related to the lifetime of a resonance 
\cite{hadrontimedelay1,hadrontimedelay2}. 
With the phase shift in general being given as 
$\delta_l \propto E^{l+1/2}$, the energy derivative 
$d\delta_l/dE \propto E^{l - 1/2}$ and leads to a singularity 
near threshold for $l = 0$. 
This problem can be overcome if instead of Wigner's time delay one considers 
the ``dwell time delay" which is a closely related concept \cite{meprl}. 
In a tunneling problem in 1-dimension for example, one finds, 
\begin{equation}\label{win9new}
\tau_{\phi}(E) \,=\, \tau_D (E) \,-\, \hbar \,
[\Im m (R)/ k] \,\,dk / dE\,,  
\end{equation}
where the first term on the right is the dwell time or 
time spent inside the barrier and the 
second term is a self-interference term which arises due to the interference of 
the incident and reflected waves. $R$ is the reflection amplitude which in 
a scattering problem gets related to the $S$-matrix.

Starting from (\ref{win9new}), replacing $R = -S$ with 
$S$ related to the complex
transition matrix in scattering as, $S = 1 \,-\, i \mu\,k \,(t_R \,+\, it_I)/\pi$, where 
$t_R$ and $t_I$ are the real and imaginary parts of the $t$-matrix 
respectively and $\mu$ is the reduced mass of the system, one obtains the
`self-interference' term in terms of $t$ as,
\begin{equation}\label{sandr}
- \hbar [\Im m (R)/k]\, \,dk/ dE\,=\, - \hbar \,\mu\, [t_R / \pi] \,\,
dk/dE\,. 
\end{equation}
Replacing the low energy behaviour of the reflection amplitude 
$R\sim e^{i(\pi - 2ak)}$ \cite{meprl}, 
$dk/dE = \mu/\hbar^2 \,k$
and the definition of the complex scattering length
$a = a_R + i a_I$,
namely, $t(E=0) = - 2 \pi a/\mu$, we see that
$$- \hbar \,[\Im m (R) / k]\, [dk/ dE]\,\, {\buildrel k \to 0 \over \simeq}\,\, \, 
2 \,a_R\, \mu / (\hbar \, k)$$ and similarly, 
$- \hbar\, \mu\, [t_R / \pi] \,[dk / dE] \,\,\, 
{\buildrel k \to 0 \over \simeq} \,\,\, 2 \,a_R \,\mu / (\hbar \, k)$.
This indeed is also the threshold singularity present in the definition of
the time delay in $s$-wave collisions near threshold.
The real scattering phase
shift for $s$-waves, $\delta \rightarrow k \,a_R$ close to threshold and
Wigner's time delay, 
$${\tau}_{\phi}(E) \,= \,2 \,\hbar\, [d\delta / dE] 
\,\,\,{\buildrel k \to 0 \over \simeq} \,\,\,2 \, a_R \, \mu / (\hbar k)$$
(note that the dwell time delay $\tau_D {\buildrel k \to 0 \over \simeq} 0$).

In \cite{meprl} the relation between the dwell and 
phase time delay in scattering was thus found to be 
\begin{equation}\label{fin}
\tau_D(E) \,=\, \tau_{\phi}(E)\, +\, \hbar \,
\mu\, [t_R / \pi] \,\,dk / dE\,.
\end{equation}
If one starts with the definition of phase time delay in terms of 
the $S$-matrix, $\tau_{\phi}(E) \,=\, 
\Re e [ -i \hbar (\,S^{-1} {dS / dE}\,)\,]$ \cite{smith} 
and uses the relation 
between $S$ and $t$ mentioned above, one gets  
\begin{equation}\label{tdsmat}
\tau_{\phi}(E) 
\,=\, {2 \hbar \over A}\,\, \biggl[\,{-\mu \over 2 \pi}\,k\,{dt_R \over dE} \,-\, {\mu^2 \,k^2 \over 
2 \pi^2}\, \biggl (\, t_I\,{dt_R \over dE}\, -\, t_R\,{dt_I \over dE}\,\biggr) \, -\, 
{\mu \over 2 \pi}\,t_R \,{dk\over dE}\,\biggr]\,,
\end{equation}
with $A = 1 \,+\, (2\mu k t_I / \pi)\,+\, (\mu^2 \,k^2 (t_R^2\, + \,t_I^2)/\pi^2) $. For elastic scattering in the absence of inelasticities,
the factor $A = 1$. Once the $t$-matrix is known, Eq.~(\ref{fin}) can be used 
to evaluate the dwell time delay in elastic scattering. 
In \cite{ourtimedelay, meprl} the above delay times for $\eta$-nucleus elastic 
scattering were evaluated with the objective of locating quasibound states 
of $\eta$ mesons and light nuclei. The $t$-matrix for $\eta d$, $\eta$-$^3$He 
and $\eta$-$^4$He elastic scattering was constructed using few body 
equations within the finite rank approximation explained in the previous 
subsection. In Figure 6 we see one such plot for the time delay in the 
reaction $\eta \, ^3$He $\to \eta \, ^3$He. As discussed above the phase time 
delay consists of a sharp singularity near threshold. 
\begin{figure}[h]
\centering
\includegraphics[width=6cm,height=6cm]{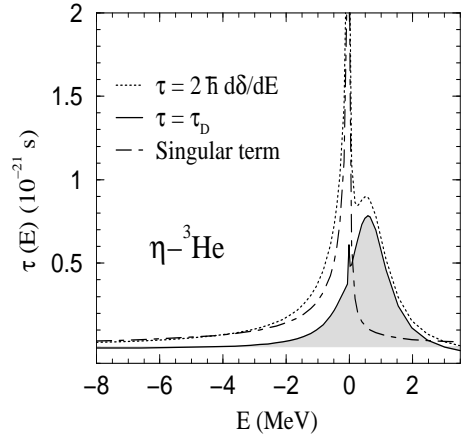}
\caption{\label{fig:eps2}Time delay in elastic $\eta$-$^3$He scattering 
\cite{meprl}. }
\end{figure}
\begin{table}
\centering
\caption{Pole values of eta-mesic light nuclear states}
\begin{tabular}{|l|l|l|l|}
\hline
 & Complex Pole (E,$\Gamma$/2) & State & Ref. \\ 
        &  in (MeV)                        &       &           \\ \hline
 $\eta d$ & -2 - $i$ 10  & Quasibound & \cite{ueda}\\
 &-$i$ 10.317  &Quasibound & \cite{rakit2} \\
 &28.06 - $i$ 24.976& Resonance & \cite{rakit2} \\
 &8.24 - $i$ 4.575& Resonance & \cite{rakit2} \\
 &3.73 - $i$ 3.405 & Resonance & \cite{rakit2} \\
 & $i$ 0.743 & Quasivirtual & \cite{wycech1} \\ 
&-24 + $i$ 27.93 & Quasivirtual & \cite{garci1}\\ 
&-0.87 + $i$ 0.95 & Quasivirtual & \cite{garci1}\\ 
& -17.1 - $i$ 17.5& Quasibound & Missed in \cite{garci1}\\
 & & & noted in \cite{ourtimedelay}\\
 &-15 - $i$ 20 &Quasibound & \cite{ourtimedelay} \\ \hline
$\eta$-$^3$He&7.03-$i$ 13.1& Resonance& \cite{rakit2} \\
&-$i$ 11.15&Quasibound & \cite{rakit2}\\
&0.5 - $i$ 0.65& Resonance& \cite{meprl} \\
&-5 - $i$ 8, - $i$ 1.95 &Quuasibound& \cite{meprl} \\ \hline 
$\eta$-$^4$He& -4.44 -$i$ 6.37, -$i$5.725 &Quasibound& \cite{rakit2} \\
&-2 - $i$ 1.75& Quasibound& \cite{meprl} \\ 
&&& \\ \hline
\end{tabular}
\end{table}
However, after subtracting 
the singular term one obtains the dwell time delay (Eq.~(\ref{fin})) which 
appears very clearly with a typical Lorentzian form of a resonance. The 
curves shown in Figure 6 correspond to an $\eta N$ scattering length input of 
$a_{\eta N}$ = 0.88 + $i$ 0.41 fm. Quasibound states in the $\eta d$ and 
$\eta$-He systems were located by varying the strength of the $\eta N$ 
interaction \cite{ourtimedelay}. 
The authors found that small $\eta N$ scattering lengths were 
more favourable for the generation of quasibound states. An example of the 
$\eta$-$^4$He quasibound system with an input $\eta N$ scattering length of 
0.28 + $i$ 0.19 fm is shown in Figure 2 in \cite{meprl}. 
With $a_{\eta N}$ = 0.88 + $i$ 0.41 fm, 
a negative time delay peak is obtained. This could correspond to a 
possible quasivirtual state centered near zero energy as discussed in 
\cite{ourtimedelay}. 
A detailed account of these results can 
be found in \cite{ourtimedelay, meprl}. 

Before ending this section we list in Tables 1 and 2 the pole positions 
for light and heavy eta mesic states found in literature. In case of heavy 
nuclei one expects stronger attraction and all the listed states are predictions 
for quasibound states. The status of the light eta-mesic nuclei is however 
different and we list the quasibound, quasivirtual and resonant states found so far. 
Since some papers list several states for varying input parameter sets, we list 
only some representative values and refer the reader to the original reference 
in the last column for all values. 
\\
\begin{longtable}{|l|l|l|}
\caption{Quasibound $\eta$-mesic states} \label{table2}
\\
\hline
Nucleus& Pole values in MeV& Ref. \\ \hline 
\endfirsthead
\hline
Nucleus& Pole values in MeV & Ref. \\  \hline
\endhead
\hline
\multicolumn{3}{|c|}{continues on the next page}\\
\hline
\endfoot
\hline
\endlastfoot
$^6$He \quad 1s&-10.7 - $i$ 7.25, -8.75 - $i$ 14.95 
& \cite{opticalthomas1}\\ \hline
$^{11}$B \quad  1s& -24.5 - $i$ 11.4, -22.9 - $i$ 23.05 
&\cite{opticalthomas1} \\ \hline
$^{12}$C \quad 1s& -1.19 - $i$ 3.67& \cite{opticalhaider} \\
& -9.71 - $i$ 17.5 & \cite{opticaloset2} \\
&- 5 - $i$ 8, -6 -$i$16 &\cite{opticaloset1} \\ \hline
$^{16}$O \quad 1s& -3.45 - $i$ 5.38 & \cite{opticalhaider} \\
&-32.6 - $i$ 13.35, -31.2 - $i$ 26.95& \cite{opticalthomas1} \\
\quad \quad \quad 1p&-7.72 - $i$ 9.15, -5.25 - $i$ 19.1& \cite{opticalthomas1} \\ \hline
$^{24}$Mg \quad 1s& -12.57 - $i$ 16.7 & \cite{opticaloset2} \\ \hline
$^{26}$Mg \quad 1s& -6.39 - $i$ 6.6 & \cite{opticalhaider}\\
&-38.8 - $i$ 14.25, -37.6 - $i$ 28.65 & \cite{opticalthomas1} \\ \hline
$^{27}$Al \quad 1s& -16.65 - $i$ 17.98 & \cite{opticaloset2}\\
\quad \quad \quad 1p& -2.9 - $i$ 20.47 & \cite{opticaloset2} \\ \hline
$^{28}$Si \quad 1s& -16.78 - $i$ 17.93 & \cite{opticaloset2} \\
\quad \quad \quad 1p&-3.32 - $i$ 20.35 & \cite{opticaloset2} \\ \hline
$^{40}$Ca \quad 1s& -8.91 - $i$ 6.8 & \cite{opticalhaider} \\
& -14 - $i$ 43, -18 - $i$ 21, -14 - $i$ 11.5& \cite{opticaloset1} \\
&-46 - $i$ 15.85, -44.8 - $i$31.8 & \cite{opticalthomas1} \\
&-17.88 - $i$ 17.19& \cite{opticaloset2} \\
\quad \quad \quad 1p& -3 - $i$ 16.5& \cite{opticaloset1} \\
&-7.04 - $i$ 19.3& \cite{opticaloset2} \\
&-26.8 - $i$ 13.4, -25.2 - $i$ 27.1& \cite{opticalthomas1} \\
\quad \quad \quad 2s& -4.61 - $i$ 8.85, -1.24 - $i$ 19.25& \cite{opticalthomas1} \\ \hline
$^{90}$Zr \quad 1s& -14.8 - $i$ 8.87& \cite{opticalhaider} \\
&-52.9 - $i$ 16.6, -51.8 - $i$ 33.2& \cite{opticalthomas1} \\
\quad \quad \quad 1p& -4.75 - $i$ 6.7& \cite{opticalhaider} \\
&-40 - $i$ 15.25, -38.8 - $i$ 30.6& \cite{opticalthomas1} \\ 
\quad \quad \quad 2s& -21.7 - $i$ 13.05, -19.9 - $i$ 26.55& \cite{opticalthomas1} \\ \hline
$^{208}$Pb \quad 1s& -18.46 - $i$ 10.11 & \cite{opticalhaider} \\
& -25 - $i$ 47, -27 - $i$ 23.5, -22 - $i$ 12.5 & \cite{opticaloset1}\\
&-21.25 - $i$ 15.88& \cite{opticaloset2} \\
&-56.3 - $i$ 16.6, -55.3 - $i$ 33.1 & \cite{opticalthomas1} \\
\quad \quad \quad 1p& -12.28 - $i$ 9.28& \cite{opticalhaider} \\
&-18 - $i$ 45, -21 - $i$ 22, -16 - $i$ 12& \cite{opticaloset1} \\
& -17.19 - $i$ 16.58& \cite{opticaloset2} \\
&-48.3 - $i$ 15.9, -47.3 - $i$ 31.75  & \cite{opticalthomas1} \\
\quad \quad \quad 2s& -2.37 - $i$ 5.82& \cite{opticalhaider} \\
&-10 - $i$ 20, -6 - $i$ 10.5& \cite{opticaloset1} \\
&-10.43 - $i$ 17.99 & \cite{opticaloset2} \\
&-35.9 - $i$ 14.8, -34.7 - $i$ 29.75 & \cite{opticalthomas1} \\
\quad \quad \quad 1d& -3.99 - $i$ 6.9& \cite{opticalhaider} \\
&-12.29 - $i$ 17.74& \cite{opticaloset2} \\
\quad \quad \quad 1f& -6.64 - $i$19.59 & \cite{opticaloset2} \\
\quad \quad \quad 2p& -3.79 - $i$ 19.99 & \cite{opticaloset2} \\
\quad \quad \quad 1g& -0.33 - $i$ 22.45& \cite{opticaloset2}\\
\hline 
\end{longtable}

\section{Reaction mechanisms for meson production}
With the momentum transfer in meson producing reactions being large,
there exist certain common features regarding the role of the
reaction mechanisms producing pions, eta mesons and kaons.
In the sections which follow, we shall discuss the cross section features 
of proton induced 
$\eta$ meson producing reactions and theoretical works which try to 
explain them. These include the $p d \to$ $p d \eta$, $p d \to \, ^3$He $\eta$ 
and the $p \,^6$Li $\to \, ^7$Be $\eta$ reactions. The latter can indeed be 
modelled in terms of the $p d \to \, ^3$He $\eta$ reaction within a cluster 
model for the nuclei $^6$Li and $^7$Be. In view of the above, 
let us start the discussion with the possible reaction mechanisms for the 
production of $\eta$ mesons in $p d$ collisions.
\subsection{One, two and three body mechanisms}\label{RM1}
\begin{figure}[h] 
\includegraphics[width=6cm,height=9cm]{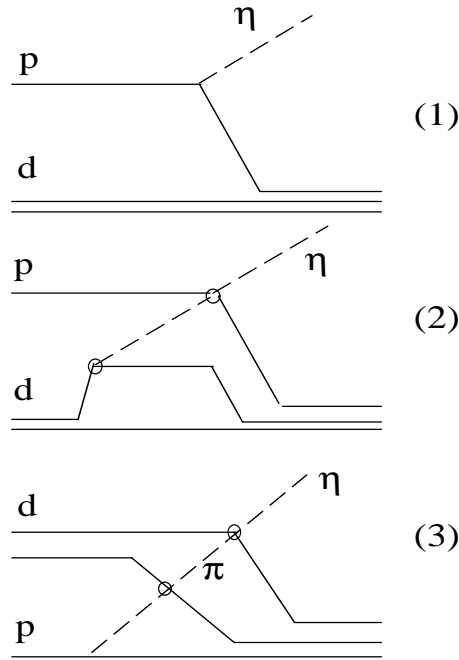}
\centering
\caption{The one-, two- and three-body meson exchange 
graphs for $\eta$ production in $p d$ collisions. Exchange graphs arising due to 
antisymmetrization are not shown here but can be found in 
\cite{lagetmechanisms}.}\label{lagraph}
\end{figure}

The need for two- and three-body mechanisms apart from the one-body mechanism 
for meson production was noticed by Laget and Lecolley \cite{LLold}. The 
three-body mechanism in particular is necessary to recover the agreement 
between theory and experiment for reactions involving $\eta$ meson 
production. Due to the large mass of the $\eta$ (547.85 MeV), the momentum 
transfer to the residual nucleus is large and is more likely shared by 
three rather than two nucleons. The one-, two- and three-nucleon graphs for 
meson production in $p d$ collisions as discussed in 
\cite{lagetmechanisms, LLold} are shown in Figure~\ref{lagraph}.
Note that in the three-body mechanism, the proton interacts in the first step 
with a nucleon in the deuteron to produce an off-shell meson which in the 
next step interacts with the other nucleon in the deuteron to produce the 
on-shell meson in the final state. 

Laget and Lecolley found that \cite{lagetmechanisms} the one- and two-body 
mechanisms in the $p d \to \, ^3$He $\eta$ reaction underestimated the 
experimental cross sections by two orders of magnitude. The role played by 
this mechanism and its limitations were later discussed in \cite{kanchan1}. 
Though a two-step model (involving the three-body mechanism) did succeed 
in reproducing the right order of magnitude of cross sections, the forward 
peaking in the angular distributions at high energies could not be reproduced 
very well. A similar model was used in \cite{kondratyuk} to study the 
$p d \to \, ^3_{\Lambda}$H K$^+$ reaction up to beam energies of 3 GeV. 
The authors found the one- and two-body mechanisms to contribute 2-3 
orders of magnitude lesser than the three-body mechanism and produced 
backward peaked angular distributions (consistent with the findings of 
\cite{kanchan1}). Similar problems in reproducing the 
angular distributions for the $p$ $d$ $\to$ $^3$He $\omega$ reaction  also have been 
reported in Refs~\cite{kondratyuk2,kanchanwasa}. A comparison of the 
one and two step processes in the $^{12}C(p,\eta)^{13}N$ reaction can be 
found in \cite{krip2}. 

Finding a mechanism which produces the right order of magnitude of the 
cross sections as well as the right peaking in the angular distributions 
remains to be an open question. In subsections 4.2 and 4.3 respectively 
we shall discuss the threshold and high energy proton 
induced $\eta$ production. The theoretical models in 4.2 are two step models 
whereas 4.3 discusses a meson exchange model which in spite of being a one 
step model seems to reproduce the data on $p d \to \, ^3$He $\eta$ 
at high energies well \cite{santraBK2}. The reason 
behind this could be the coherent sum of meson exchange diagrams which is 
not used in the other approaches or the difference in the nuclear wave 
functions used. A calculation based on the meson exchange model and 
using a rigorous few body formalism for the inclusion of the final state 
eta-nucleus interaction could turn out to be useful in understanding 
the reaction mechanisms involved in the $\eta$ producing reactions.  

\subsection{Proton induced $\eta$ production on light nuclei 
and cross section features near threshold}\label{RM2}
The cross sections of reactions such as the $p$~$d$~$\to$~$^3$He~$\eta$,  
$p$~$d$~$\to$~$p$~$d$~$\eta$ and $p$~$n$~$\to$~$d$~$\eta$, 
display very similar features
as seen in 
Figs~\ref{fsq} and \ref{totxn}. 
We saw earlier that the sharp rise near threshold 
is attributed in literature to the strong $\eta$-nucleus FSI. 
The production mechanisms for these reactions 
are based on diagrams discussed in the previous 
subsection. However, apart from such descriptions, there also 
exist meson exchange models which are used to describe the 
$\eta$ production in $p N$ collisions \cite{lagetppeta,santraBK1,rshyamppeta, 
Teresa,kanzo1,kanzo2,deloff_pp,kondratyuk3} and then further applied 
to $\eta$ production with $p$-nucleus collisons. Such models will be 
discussed in the next subsection. 
   
In the present subsection, 
as an example, we briefly discuss the formalism where 
a two step model is used to produce the $\eta$-$^3$He system 
in the $p$~$d$ collision and where few body equations are solved to include the final state interactions. The same formalism can be used to study the 
 proton (or deuteron) induced $\eta$ production on other light nuclei 
\cite{wilkin3,neelam1,neelam2,khalili,neelam7Be,ulla}. 
The $p$~$d$~$\to$~$p$~$d$~$\eta$ reaction, where, once again the two step model 
has been used, will be discussed in one of the subsequent sections. 

The Born amplitude for the two step 
process, $< |T_{p \,d \rightarrow \,^3{\rm He}\, \eta}| >$, can be written 
as \cite{lagetmechanisms,wilkin2,kondratyuk,kanchan3,LLold}
\begin{eqnarray}\label{born}
< |T_{pd \rightarrow ^3{\rm He}\,\eta}| >=i \int {d\vec{p_1}\over (2\pi)^3}
{d\vec{p_2}\over (2\pi)^3} \sum_{int\,m's} <p n \,|\,d>
\,<\pi\,d |T_{pp\, \rightarrow\, \pi\, d}| p\,p>
\\ \nonumber
 \times{1\over (k_\pi^2-m_\pi^2+i\epsilon)}
\, <\eta\,p \,|\,T_{\pi N \rightarrow \eta p}\,| \pi\,N>
\,\,<\,^3{\rm He}\,|\,p\,d>\,,
\end{eqnarray} 
where the sum runs over the spin projections of the intermediate
off-shell particles and $k_{\pi}$ is the four momentum of the intermediate pion.
The matrix elements $<p n|d>$, $<^3$He$|p d>$,  $<\pi\,d |T_{pp\, \rightarrow\, \pi\, d}| p\,p>$ and
$<\eta\,p \,|\,T_{\pi N \rightarrow \eta p}\,| \pi\,N>$  can be obtained 
from Refs~\cite{bhalerao,etaN5,paris,he3,pppid}, for example. 
The FSI can be incorporated in the formalism by writing the $\eta-^3$He wave function as a sum of the plane wave and the scattered
wave as
\begin{equation}\label{wave}         
<\,\Psi_{\eta \,^3{\rm He}}^- \,| \, =\, <\, \vec{k_{\eta}}\,| \, + \,
\int { d\vec {q} \over (2 \pi)^3 } \, {<\,\vec{k_{\eta}} \,|\,
T_{\eta \,^3{\rm He}}\, | \, \, \vec{q}\,>
\over E(k_{\eta}) \, - E(q)\, + \,i\epsilon} \, <\vec{q}\,|,
\end{equation}
where $T_{\eta\,^3{\rm He}}$ is the T-matrix for $\eta\,^3$He elastic
scattering.
In this way, the $T$-matrix  for the $p$~$d$~$\to$~$^3$He~$\eta$ process, including the $\eta\,^3$He interaction, becomes
\begin{eqnarray}\label{tmat2}
&&T =\, <\,\vec{k_\eta}\, ; \, m_3\,|\, T_{p d \rightarrow \,^3{\rm He}
\,\eta}\,|
\,\vec{k_p}\, ; \, m_1 \, m_2\,> + \\ \nonumber
&&\sum_{m_3^\prime} \int { d\vec {q} \over (2\pi)^3} {<\, \vec{k_\eta}\,
; \, m_3\,|         
\, T_{\eta\, ^3{\rm He}}\, |\,\, \vec{q}\, ; \, m_3^\prime\,> \over E(k_\eta)\,
- \,E(q)\,
+\, i\epsilon}\,
<\vec{q}\, ; \, m_3^\prime\,| T_{p d \rightarrow \,^3{\rm He} \,\eta}\,|
\,\vec{k_p}\, ; \, m_1 \, m_2>.
\end{eqnarray}

The $\eta$-light nucleus system constitutes of few particles and hence the corresponding interaction
 can be obtained accurately by solving few body equations. A more detailed discussion on the 
 importance of solving few-body equations for light systems will be made in Section 5.1.
But, in brief, it can be mentioned at this point that the few body equations take into account the multiple 
off-shell scattering of the $\eta$-meson on different nucleons which are bound  as a nucleus. Indeed, 
the $\eta ^3$He interaction was obtained by solving few body equations by considering the $^3$He nucleus 
to remain in its ground state in 
Refs~\cite{rakit1,rakit2,rakit3,kanchan1,kanchan2,kanchan3} and the resulting
 amplitude was used to calculate  Eq.~(\ref{tmat2}) in 
Refs~\cite{kanchan1,kanchan2,kanchan3}.  
 \begin{figure}[h!]
\centering
\includegraphics[width=15cm]{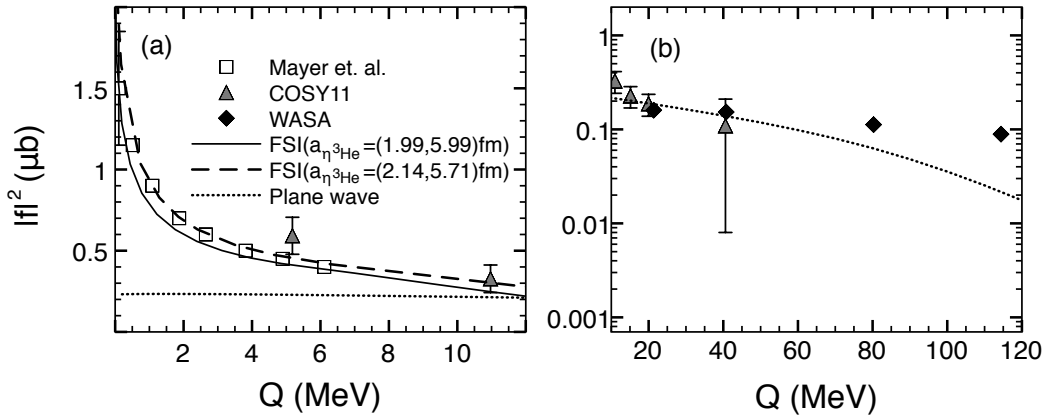}
\caption{The squared amplitude of the $p$~$d$~$\to$~$^3$He~$\eta$ reaction for the excitation energies, 
$Q(=\sqrt{s} -M_{^3{\textrm He}}-M_\eta) < 12$ MeV 
is shown in the left  panel and for $10 < Q < 120$ is shown 
in the right panel. The figure shows data from Refs~\cite{dataMayer,pd3He2,pd3He4} as empty squares, filled 
triangles and filled rhombuses respectively. 
The meaning of the lines \cite{kanchan1,kanchan2,kanchan3} 
is explained in detail in the text.}\label{fsq_pd}
\end{figure}

The squared amplitude for the $p$~$d$~$\to$~$^3$He~$\eta$ reaction calculated within such a formalism is
 shown in Figure~\ref{fsq_pd}.
The  squared amplitude obtained by assuming plane waves for the final state particles  is shown by dotted lines in
 Figure~\ref{fsq_pd}. The figure shows that these results 
cannot explain the sharp structure present in the near-threshold data of 
Refs~\cite{dataMayer} (shown by empty squares) and 
\cite{pd3He2} (shown by filled triangles). 
However, these data in the threshold region can be well reproduced 
 by taking the $\eta-^3$He FSI into account as shown by the solid and dashed lines in Figure 8(a). These results
  have been obtained by incorporating the FSI calculated by solving few-body equations taking two different input  
  $\eta N$ interactions corresponding to scattering lengths $a_{\eta N} = (0.75,0.27)$ fm \cite{fix,green} and 
  $(0.88, 0.41)$ fm \cite{garci2}, which give the 
$\eta ^3$He scattering lengths to be 
$a_{\eta^3{\textrm He}} = (1.99,5.99)$ fm and $(2.14,5. 71)$ fm, 
respectively (for more details on the calculations which lead to the 
results shown by solid and dashed lines, 
please look at Refs~\cite{kanchan1,kanchan2,kanchan3}).

Figure 8(b) shows the data \cite{pd3He2} and the 
plane wave calculation done in
  Ref.~\cite{kanchan2,kanchan3} for the squared amplitude of 
the $p$~$d$~$\to$~$^3$He~$\eta$ reaction at 
   higher excitation energies, $Q > 10$ MeV. 
It can be seen that the calculation done by assuming plane waves 
   for the final state can explain the data up to $Q \sim 60$ MeV but 
starts deviating from the data beyond that. 
   It is possible that a different production mechanism dominates in this energy region. It is important to mention 
   one of the shortcomings of the two-step model here. This model leads to isotropic angular distributions very 
   close to the threshold region but one obtains backward peaked angular distributions at higher excitation energies.   

\begin{figure}[h!]
\centering
\includegraphics[width=8cm]{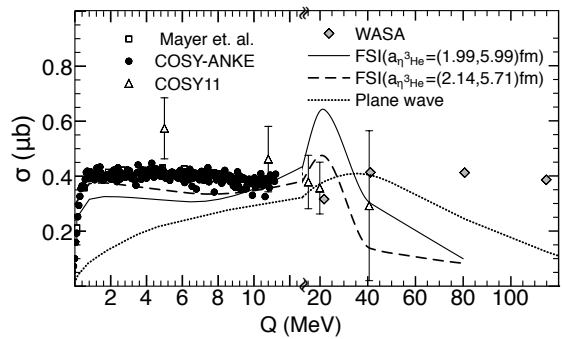}
\caption{The total cross sections for the $p$~$d$~$\to$~$^3$He~$\eta$ reaction. The data from
 Refs~\cite{dataMayer,dataMersmann,pd3He2,pd3He4} are shown, respectively, by empty squares,
  filled circles, empty triangles and empty rhombuses. The theoretical 
curves \cite{kanchan2} are discussed in detail in the text. }\label{totxn2}
\end{figure}

The  total cross sections for the $p d \rightarrow \,^3$He $\eta$ reaction can be calculated as
\begin{equation} \label{crossangular}
\biggl ({d\sigma \over d\Omega}\biggr )_{c.m} 
= {1 \over 12} {m_p\, m_d \,m_{3} \over (2 \pi E_c)^2} \,{p_c^f \over p_c^i}\, \sum_{if}\, |T_{fi}|^2,
\end{equation} 
where $m_p$, $m_d$, $m_{^3{\textrm He}}$ are the masses of the proton, deuteron and the $^3$He nucleus, 
and $E_c$, $p_c^f$, $p_c^i$ refer to the total center of mass energy and momenta in the final and initial state, respectively.
We show the data on the total cross section and the results of its calculations done by using the two-step model for the 
production mechanism and a solution of the few-body equations for the 
$\eta \,^3$He FSI (as 
done in Refs~\cite{kanchan1,kanchan2,kanchan3}) in Figure~\ref{totxn2}. In this figure too, the dotted, dashed and solid lines,
 respectively, show the results of the calculations done in the plane wave approximation and by taking the FSI into account 
 using two different inputs, as explained above.  Once again, the near threshold data seems to get well explained by this 
 model but the agreement between the two becomes poor at relatively higher energies. 
\subsection{Meson exchange model for eta production at high energies}\label{RM3}
As discussed above, 
the two-step model describes well the available eta meson 
production data on the $p d\rightarrow \,^3$He $\eta$ 
reaction near threshold. The angular distributions near threshold are 
isotropic. At high energies, though the model still reproduces the right 
order of magnitude of the cross sections \cite{lagetmechanisms, kondratyuk,
kanchan1}, it fails to reproduce the forward 
peaks in the angular distributions. In \cite{kanchan1} the authors discussed 
the limitations of this model in detail and showed that only some ad hoc 
manipulations of the model such as restricting the intermediate pion to 
be on shell and to be produced in the forward direction could change the 
backward peaked $\eta$ angular distributions in this model to forward ones.
This discrepancy in the angular distributions gives rise to the need for 
investigating models other than the two step model which has been widely
used in literature.

In what follows, we shall discuss a model for the $p d \to$ $\,^3$He $\eta$ 
reaction where the $\eta$ is produced via the $p N \to$ $p N \eta$ reaction. 
The elementary $p N \to$ $p N \eta$ reaction is described within a 
boson exchange model (BEM) involving the exchange of the $\pi$, $\eta$, 
$\rho$ and $\omega$ mesons. 
The calculated cross sections over a wide energy range for the 
$p p \rightarrow p p \eta$ reaction have been found in this model 
to agree very well with the measured ones. 
In the following, therefore, we first give a brief description of the 
existing BEM descriptions of  
the $p p \rightarrow p p \eta$ reaction in literature 
and then discuss the work which uses the BEM for the $p N \to p N \eta$ 
reaction as an input to describe the $p d \rightarrow \,^3$He $\eta$ 
reaction. 

\subsubsection{$pp\rightarrow pp\eta $ reaction} 
BEM based detailed studies of the $p$~$p$~$\rightarrow$~$p$~$p$~$\eta$ reaction can been found in 
Refs~\cite{lagetppeta,santraBK1,rshyamppeta,kanzo1,kanzo2,deloff_pp}. All these calculations recognize that the $\eta N$ interaction is 
dominated by the $S_{11}(1535)$ nucleon resonance. Hence, all of them  assume that 
in the $pp$ collision in the incident channel one of the protons (projectile or the target) gets excited to the 
$S_{11}(1535)$ resonance, which then decays to $\eta p$, 
i.e. $p p \rightarrow NS_{11}(1535)\rightarrow p p \eta $. The t-matrices corresponding to the projectile and target 
excitations are called respectively  ``direct" and ``exchange" 
terms. In the full t-matrix for the process $ p p \to p p \eta$,   
these t-matrices  appear as a coherent sum with opposite signs because of the antisymmetry of the proton-proton wave functions, i.e.
\begin{equation}
T(p p \rightarrow p p \eta )=T_D(p p \rightarrow p p \eta )-T_E(p p \rightarrow p p \eta ).
\end{equation}
The choice of the exchanged mesons between the interacting protons in the entrance channel is also guided, 
to a certain extent, by the dominant decay modes of the $S_{11}$ resonance. All the calculations thus have 
a pion-exchange because the $S_{11}$ resonance  strongly couples to a pion in addition to an eta meson. 
Coupling to a $\rho $ meson is included because of the large radiative width of the $S_{11}$ resonance and 
the vector meson dominance (VMD) in the radiative coupling. Thus all the these studies of the elementary reaction, 
$pp\rightarrow pp\eta $ include $\pi +\rho $-exchange. Refs \cite{santraBK1, rshyamppeta}, however, include the exchange of 
some other mesons too. However, they find that the $\pi +\rho $-exchange plays the deciding role. The parameters 
associated with the coupling and propagation of all these mesons are decided in all the calculations by data 
on independent  relevant processes. Thus, in that sense, the calculations were parameter free. All other couplings being similar, 
the $\rho NN^*$  coupling in \cite {rshyamppeta} is different by being $\gamma _5\sigma _{\mu \nu }$  from 
those in the other two studies \cite{lagetppeta, santraBK1}, where it is $\gamma _5\gamma _\mu $. This difference, as we will see later, 
leads to very different magnitude of the $\rho $-exchange contribution to cross sections. 

The above calculations also included the final state interaction 
(FSI) amongst $p, p$ and $\eta $ in different ways and to varying degrees. 
The authors in \cite{lagetppeta} incorporate the FSI by modifying the $^1S_0$ $ pp$ wave only.
 For this they use the Paris potential at all the  
energies except near threshold, where they use the Coulomb corrected effective range expansion. These calculations do not 
include the interaction of the $\eta $ meson with the protons. The calculated total cross sections are presented up to 
3 GeV beam energies. They reproduce the available data, which were very limited, at the time of this calculation. This calculation also found that 
the $\rho $-exchange transition amplitude dominates. 

In \cite{rshyamppeta} the  FSI is included amongst all three outgoing 
particles, taking them pairwise, within the Watson-Migdal theory. Here the t-matrix without FSI is multiplied by a 
sum of three two-body subsystem factors, each one of them described through 
the Jost function written in terms of the corresponding 
effective range expansion parameters and corrected for the 
Coulomb interaction in the $pp$ pair. The FSI is included only for 
s-wave. The calculated total cross sections agree well with the measured values over a large energy range. The author also finds, 
in contrast to the findings in \cite{lagetppeta, santraBK1}, that the $\rho $-exchange term does not dominate the cross sections. 
This observation seems to be supported by the recent analyzing power measurements on the $pp\rightarrow pp\eta $ reaction. 

The treatment of the FSI in \cite{santraBK1} follows a different approach, 
given earlier by the same  authors in the study of the  
$p p \rightarrow n \Delta ^{++}$ reaction \cite{BKsantra}. The FSI amongst $pp\eta $ is envisaged to consist of the interaction 
between $\eta $ and a proton, and that of this pair and another proton. Since the dominant effect of the former is to excite 
the $S_{11}(1535)$ resonance, the t-matrix for the $p p \rightarrow p p \eta$ process  is decomposed into a t-matrix for the transition 
$pp\rightarrow pS_{11}(1535)$ and the decay probability of $S_{11}$ to $\eta p$. The interaction between $\eta p$ and the other proton 
is incorporated  using a distorted wave for the $pS_{11}$ system in the transition matrix. The initial state interaction is also 
included  in this work by using a distorted wave for the $pp$ wave in the initial state (for details see \cite{santraBK1}). Calculated 
total cross sections in this work are found to agree very well with the measured ones from threshold to high energies. Like in 
\cite{lagetppeta} the dominant contribution to the cross section is found to come from the $\rho$-exchange.

\subsubsection{$p d \rightarrow \,^3$He $\eta$}
Since the BEM seems to be successful in reproducing the elementary $\eta$ 
production data in $pp$ collisions, it is worth reviewing a model for 
the $p d \to \,^3$He $\eta$ reaction based on this input \cite{santraBK2}.  
Though this model appears at a first glance to be quite similar to the one 
step model discussed in \cite{lagetmechanisms,komarov}, it differs in 
details and the authors reproduce the right order of magnitude 
of the $p d \to \,^3$He $\eta$ cross sections in addition to the forward
peaked angular distributions. Let us briefly review this work first and then
compare it with other works in literature. 

\begin{figure}[h!]
\centering
\includegraphics[width=12cm]{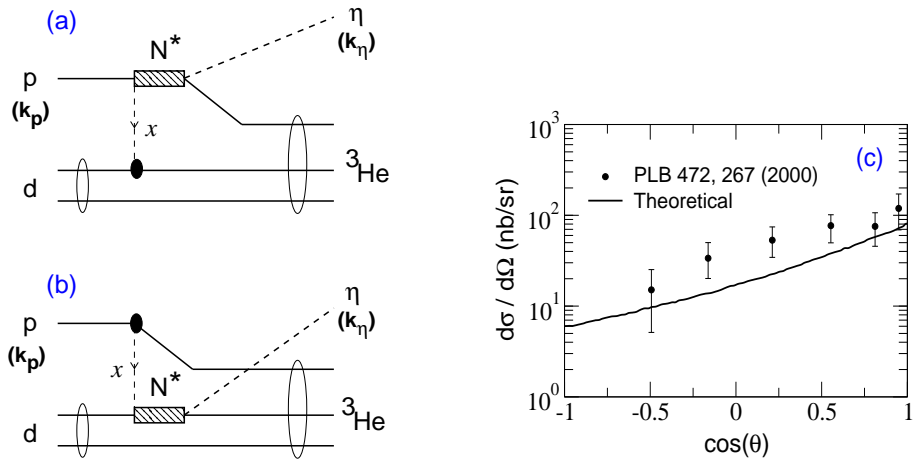}
\caption{Schematic diagram for the $p d \rightarrow \,^3$He $\eta$ 
reaction with (a) projectile excitation and (b) target excitation. $x$ 
denotes the exchanged meson which can be $\pi$, $\eta$, $\rho$ or $\omega$ 
meson. In (c) is shown the  
angular distribution of the $\eta $ meson at a beam 
energy of 88.5 MeV above threshold. The data are from
 Ref.~\cite{pd3He5} and the solid line from \cite{santraBK2}.} 
\label{dig}
\end{figure}

The authors in \cite{santraBK2} consider 
two diagrams corresponding to the excitation of 
the projectile proton and a nucleon in the target nucleus $^3$He. 
The transition matrix for a typical diagram, Figure~\ref{dig}(a) is written as
\begin{eqnarray}
\nonumber
T_{fi} = & &\frac{1}{(2\pi )^6}\int d\vec{k_1}d\vec{k_2}\psi_{He}
(\vec{k_1},\vec{k_2})\Gamma_ {NN^*\eta}(k_\eta)\, \\ \nonumber
& & \times 
G_{N^*}(\vec{Q})V_{NN\rightarrow NN^*}(\vec{q}, 
\omega )\psi _d(\vec{K}),
\end{eqnarray}
where $\vec{Q}=2\vec{k_\eta }/3+\vec{k_1}$, $\vec{q}=\vec{k_p}-2\vec{k_\eta}/3$, and 
$\vec{K}=\vec{k_\eta}/3-\vec{k_p}/2+\vec{k_1}+\vec{k_2}$. 
$N^*$ denotes the $S_{11}(1535)$ resonance. $G_{N^*}$ and $\Gamma _{NN^*\eta}(k_\eta)$ 
denote its propagation and decay vertex into $N\eta $ respectively. 
The transition potential $V_{NN\rightarrow NN^*}(\vec{q}, \omega)$
is written in terms of the vertex functions at the $x NN$ and $x N N^*$ 
vertices ($x = \pi, \eta, \rho$ or $\omega$ meson) and the meson propagator.
The bound state deuteron and $^3$He wave functions are described by the 
Hulthen and Gaussian wave functions respectively (see \cite{santraBK2} for
details) 
and consist only of the 
$s$-wave components. The parameters of the Gaussian wave function are 
chosen to reproduce the experimental mean square radius of $^3$He.
%
The angular distributions were evaluated 
as in Eq.~(\ref{crossangular}) using the transition amplitude discussed above. 
Calculations were done for a beam energy of 
88.5 MeV above threshold where data from the GEM collaboration at COSY exist \cite{pd3He5}. 
In these calculations the target and projectile excitations were added with opposite signs, 
and include the excitation of both, the neutron and proton in 
$^3$He. These results along with the data are shown in Figure~\ref{dig}, 
and are seen to reproduce them well. 

The total cross sections obtained by integrating Eq.~(\ref{crossangular}) 
over all angles also reproduced the data 
up to 1 GeV beam energy very well. 
The sensitivity of cross sections 
to different exchanged mesons in the transition potential was also studied. 
The authors find that though 
the $\rho $-exchange contribution dominates, 
contributions due to  exchange of $\omega $ and $\pi $ are also important.    
It may, however, be mentioned that above calculations do not include FSI 
between $\eta $ and $^3He$, whose effect could be significant near 
threshold. 

The above results \cite{santraBK2} contradict the findings of some 
earlier works in literature. In \cite{lagetmechanisms} for example, the 
authors evaluate the transition amplitude corresponding to the
diagram (2) in Figure \ref{lagraph}. Though the diagram appears similar 
to that in Figure (\ref{dig}), the transition amplitude is evaluated in a 
very different way. Apart from some Clebsch-Gordon coefficients, it is 
written in terms of the transition matrix for the $\pi^+$ d $\to p p$ reaction 
and an overlap of the deuteron and $^3$He wave functions. The authors in 
\cite{lagetmechanisms} compare their calculation with data on the $p d \to \, 
^3$He $\eta$ reaction up to 2.5 GeV, however, only at one angle, namely, 
$\theta_{\eta} = 180^o$. They mention the three body mechanism to be the 
appropriate choice since the two body graphs (see Figure 7) underestimate the data by 
two orders of magnitude. Conclusions similar to those of 
Ref.\cite{lagetmechanisms} regarding the role of the one and two step 
mechanisms,  
however for the $p d \to$ $p d \eta$ reaction 
were obtained in \cite{neelam1}. Finally we mention the findings of 
Ref. \cite{komarov} where the authors studied the 
reaction mechanisms for the $p d \to \, ^3$H$_{\Lambda}$ $K^+$ reaction. 
We mention these results since the K$^+$ is almost as heavy as the $\eta$ 
meson and hence the kinematics and the momentum transferred to the nucleus 
in this reaction must be similar to that in the $p d \to \, ^3$He $\eta$ 
reaction. The angular distributions of the $K^+$ mesons 
using the two step model are also backward peaked at high energies as in 
case of the $\eta$'s. The one step model produces forward peaked angular
distributions, however, reduces the cross sections by
2-3 orders of magnitude. 
                
In summary we can say that the various calculations in literature point 
toward some missing components in the understanding of the reaction 
mechanisms of the $p d \to \, ^3$He $\eta$ reaction. All these works 
do seem to agree on one point that the forward peaked angular distributions 
cannot be reproduced within the two step model and that the one 
step model does reproduce the forward peaking. However, they do not agree 
on the magnitude of the cross sections produced within the one step model.
Hence, it would be useful to perform 
further investigations of the $p d \to \, ^3$He $\eta$ reaction within the 
one boson exchange model using more refined 
wave functions and including the effects of the final state interaction.  
It would be useful to obtain more data on this reaction at high 
energies in future.


\section{Eta meson interaction with nuclei in the final state}
\label{FSI}

Though the main objective of the entire eta meson related program has 
been the study of the eta nucleus interaction and location of eta
mesic states, the extraction of this information from available
data is often based on approximate methods. In the subsequent sections 
we point out the importance of few body equations and their application 
to study the effects of the $\eta$-nucleus interaction in the final states 
of the $p n \to$ $d \eta$, 
$p d \to$ $p d \eta$ and the $p \,^6$Li $\to$ $^7$Be $\eta$ 
reactions. 

\subsection{Approximate methods and the need for few body equations}

As is evident from Figure 8, $\eta$-producing reaction amplitudes show a 
strong energy dependence close to threshold but seem to approach 
a constant value for large excess energies. Theoretical analyses attribute the 
sharp rise to the final state interaction 
(FSI) between the $\eta$ and the nucleus. The interaction is dominantly $s$-wave and 
its effect reduces rapidly away from threshold. 
The above features seem to make it a good candidate for the use of the 
Watson-Migdal approximation \cite{watsonbook} 
which is indeed often used (with a further simplification) 
in the analysis of eta producing reactions.
In what follows, we briefly explain this approximation and the related conclusions 
about eta-nucleus scattering lengths drawn from it. We then go over to the 
incorporation of the FSI using a few body transition matrix and point out 
the drawbacks of 
the use of an on-shell approximation (such as the one mentioned above) 
for the FSI. 
\subsubsection{Scattering length approximation} 
Within the approach of Watson and Migdal, the energy dependence of the reaction is 
determined by the on-shell scattering amplitude of the final state 
\cite{sibirtsev}, namely, 
\begin{equation}\label{SLA1}
f_{FSI} (k) = {1 \over k \cot{\delta} - i k} \, ,
\end{equation}
where $k$ for example is the center of mass momentum of the $\eta$ and the nucleus and 
$\delta$ the corresponding $s$-wave phase shift which near threshold can be approximated by the 
effective range expansion
\begin{equation}\label{effrangeform}
k \cot{\delta} = {1 \over a } + {r_0 \over 2} k^2 \, + \, ... \,, 
\end{equation}
where $a$ is the scattering length. 
If we simplify the above expression further by neglecting terms of order $k^2$ 
and higher, 
the squared reaction amplitude for an $\eta$ producing reaction 
can be written as
\begin{equation}\label{SLA2}
|f|^2\,=\,f_B\,\times\,{|a|^2 \over 1 \,+ \,2 k \Im m \,a \,+ \,|a|^2 k^ 2}
\end{equation}
where $f_B$ represents the squared amplitude in the absence of FSI. As we 
move away from threshold, the $k^2$ term in Eq.~(\ref{effrangeform})
will be negligible only if the effective range parameter $r_0$ is 
much smaller than $a$. 
The form that is more often used in the context of the eta-nucleus 
FSI is 
\begin{equation}\label{SLA3}
|f|^2 = |f_p|^2 \, \times \, {1 \over (1 \,+ \,2 k \Im m a \,+ \,|a|^2 k^ 2)} 
\end{equation}
where $|f_p|^2$ is an arbitrary factor fitted to reproduce the 
correct magnitude of the cross sections. 

In \cite{wilkinFSI}, the author reproduced the data on  
the squared amplitude $|f|^2$ for the $p d \to$ $^3$He $\eta$ reaction using 
the above approximation along with an 
eta-nucleus scattering length, $a_{\eta ^3 {\rm He}}$ =  -2.31 + $i$ 2.57 fm, 
which was obtained 
by using the lowest order optical potential for the $\eta \, ^3$He system. 
With the potential given as, 
\begin{equation}
2\, m^R_{\eta N}\, V_{opt}(r) \, =\, - 4 \pi \, A\, \rho(r) \, a_{\eta N}\, ,
\end{equation}
the phase shift could also be determined. This enabled the author to use 
the better 
approximation in Eq.~(\ref{SLA1}) rather than just the one in Eq.~(\ref{SLA3}) 
and little difference between the two results was found. However, the 
normalization factor in both cases was simply a free parameter.
\subsubsection{Half-off-shell $\eta$-nucleus T-matrix} 
In a proper description of the FSI between the $\eta$ and the nucleus, one must 
consider the fact that the $\eta$ meson can also be produced off the mass shell and 
eventually brought on-shell due to its interaction with the nucleus. 
In a few body approach, the off-shellness enters the theory by 
expressing the final state wave function as a solution of the 
Lippmann-Schwinger equation, written as
\begin{equation}
\langle \Psi^-(\vec{k})|\,=\,\langle \vec{k}|\,+\,\int {d\vec{q}\over
(2\pi)^3}\,\,{\langle \vec{k}|T|\vec{q} \rangle\over E(k)\,-\,E(q)
\,+\,i\epsilon}\,\langle \vec{q} |,
\end{equation}
where $k$ is the final state c.m. momentum. The half-off-shell 
$\eta$-A $T$-matrix, $\langle \vec{k}|T|\vec{q} \rangle$ is generated
by solving few-body equations for the $\eta$-A system. The propagator in 
the scattering term generates two terms originating from the 
principal-value and the residue parts. Physically they represent the
off-shell and the on-shell scattering between the particles in the
final state. 
We shall see below 
that the 
contribution of the rescattering of off-shell particles in the final state 
is indeed large and the off-shell term contributes much more to the FSI 
than the on-shell rescattering term. 

In Figure 11a, we compare the amplitude squared $|f|^2$ generated using 
few body equations for the $\eta ^3$He FSI \cite{kanchan3} 
with data \cite{dataMayer}. 
The elementary $\eta N$ t-matrix which is chosen to be the 
input to the few body calculation leads to an $\eta N$ scattering length of 
(0.88, 0.41) fm and the few body t-matrix produces $a_{\eta ^3 {\rm He}}$ = 
(2.41, 5.71) fm. One can see in Figure 11b that the scattering length 
approximation (SLA) of Eq.~(\ref{SLA3}) with a value of 
$a_{\eta ^3 {\rm He}}$ = (-2.31, 2.57) fm as in \cite{wilkinFSI} 
reproduces the data equally well too.  
If on the other hand, we choose to use the value of 
$a_{\eta ^3 {\rm He}}$ = (2.41,5.71) fm as obtained from the few body 
equations with the same multiplicative factor of $|f_p|^2$ = 2, the 
SLA underestimates the data. The purpose of this exercise is to emphasize 
that any conclusions about the magnitude or sign of the eta-nucleus 
scattering length based on fits to data using the SLA 
with arbitrary multiplicative factors can be quite misleading. 

\begin{figure}[h!]
\includegraphics[width=0.49\textwidth,height=6cm]{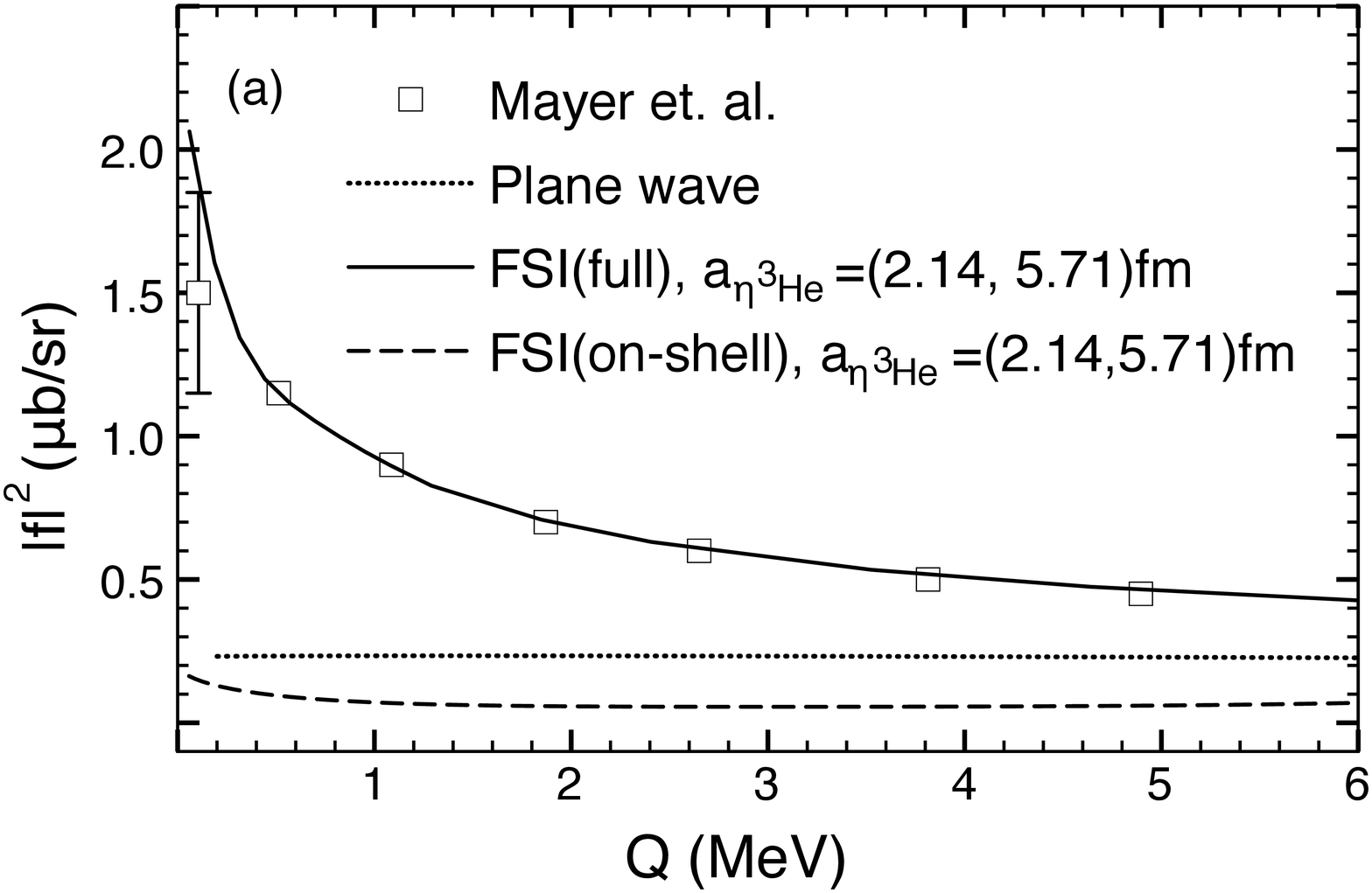}
\includegraphics[width=0.49\textwidth,height=6cm]{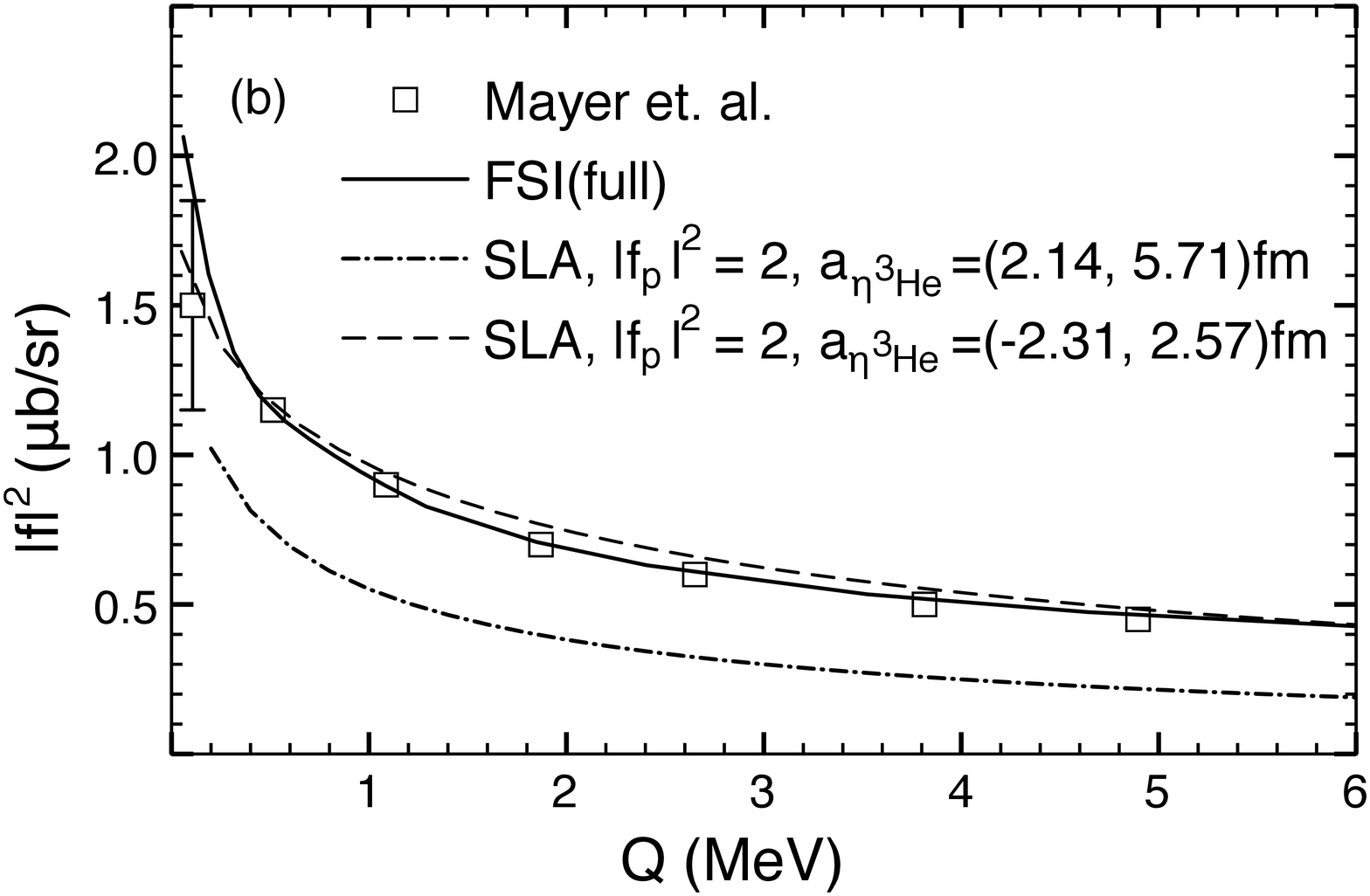}
\caption{Importance of off-shell rescattering in the $p d \to$ $^3$He $\eta$ 
reaction with (a) showing the contribution of the on-shell term in 
the few body transition matrix \cite{kanchan3} and 
(b) a comparison of the full few body calculation 
with the scattering length approximation using the $\eta ^3$He 
scattering length obtained in \cite{wilkinFSI} (dashed line) and 
the scattering length obtained from the few body t-matrix (dash-dotted line).}
\label{contri}
\end{figure}

Recognizing the significance of the few body approach, 
we present in the next sub-section a discussion
of the $p n \to$ $d \eta$, $p d \to$ $p d \eta$ and the 
$p \,^6$Li $\to$ $\eta$ $^7$Be reactions where use of few body equations 
to describe the final state interaction between $\eta$ mesons and nuclei is 
made.

\subsection{Final state $\eta$ nucleus interaction in proton induced
reactions}
We begin the discussions in this section with the simplest eta-nucleus system, 
namely, the $\eta$-deuteron system and then go over to the 
$\eta \, ^7$Be case. The $p d \to \eta \, ^3$He reaction which has been discussed 
in the earlier sections will not be dealt with over here. In what follows, we shall 
present three different approaches used to describe the $\eta d$ interaction using 
few body equations.  
\subsubsection{The $p n \to$ $d \eta$ reaction}
The cross sections for the $p n \to$ $d \eta$ reaction were calculated in a series 
of works by Garcilazo and Pe\~na \cite{Teresa,garci2005,garci2008} and various 
aspects of the production mechanism  as well as the initial and final state interaction 
were studied. The production mechanism involved the exchange of the $\pi$, $\eta$, $\sigma$ 
and heavier mesons as shown in Figure (12a). The final state $\eta d$ interaction was described 
by using Faddeev equations for $\eta$d elastic scattering (see Figure (12b)). In \cite{Teresa}, 
the authors used a non-relativistic three body model of the $\eta N N$ system where all three 
particles interacted through pairwise interactions which were represented with separable potentials. 
The orbital states between the spectator particles and the center of mass of interacting pairs were 
restricted to s-waves. The relative orbital states for the interacting pairs were taken 
as $S_{11}$ for the eta-nucleon pair and $^3S_1$ for the nucleon-nucleon pair. 
\begin{figure}[h!]
\begin{center}
\includegraphics[width=0.4\textwidth,height=2.5cm]{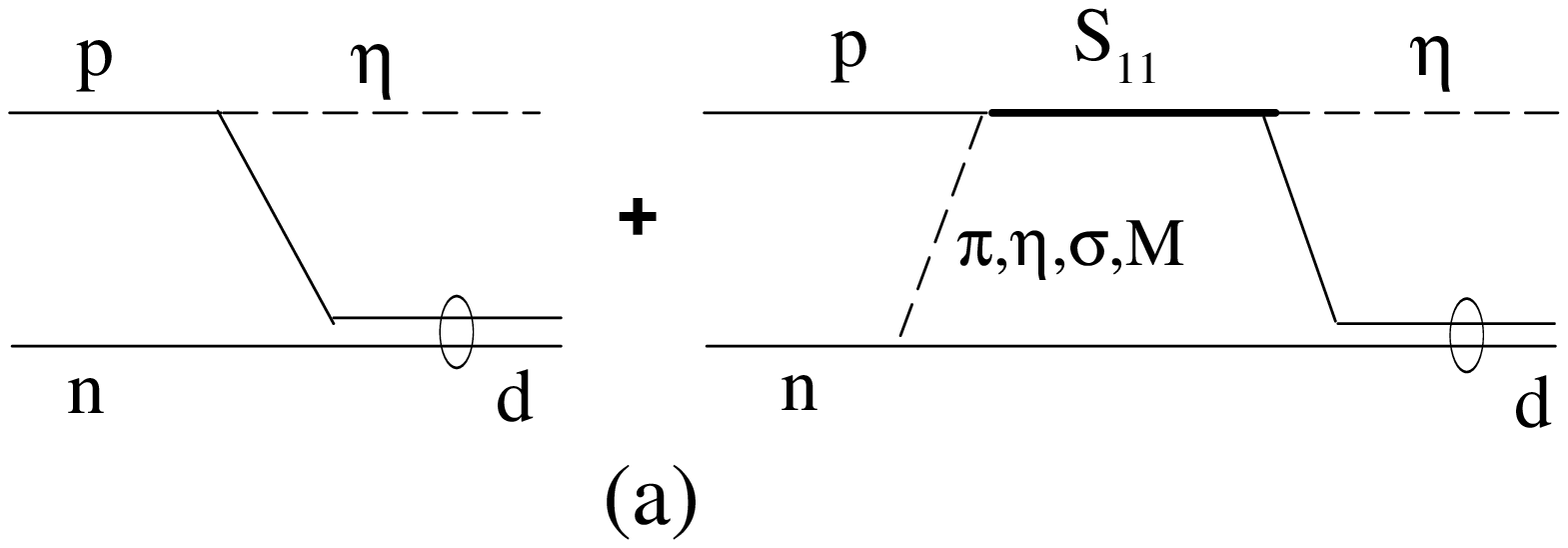}
\hskip1cm
\includegraphics[width=0.4\textwidth,height=2.5cm]{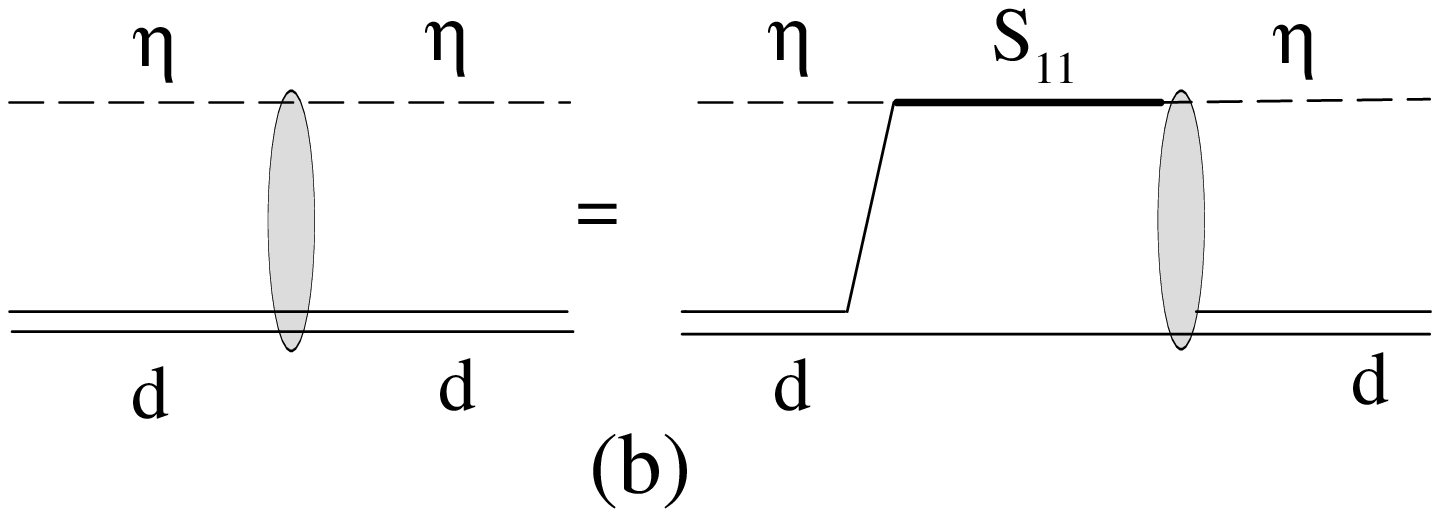}
\end{center}
\caption{Model for (a) production mechanism and (b) final state interaction 
in the $p n \to$ $d \eta$ reaction. 
  } 
\end{figure} 

The authors performed calculations using different inputs of the $\eta N$ interaction 
corresponding to a range of $\eta N$ scattering lengths. The overall agreement with data, 
including that at higher energies was found for a small value of the real 
part of the scattering 
length, namely 0.42 fm. The non-relativistic calculations in \cite{Teresa} were followed 
by a relativistic version \cite{garci2005} of the three body equations which incorporated 
relativistic kinematics and the boost of two-body meson-nucleon and nucleon-nucleon interactions. 
Though the boost effects were found to be small, relativistic effects on the range and strength of 
the pion exchange contribution to the reaction mechanism were found to be large. 
Apart from these results the authors also found that the initial state interaction, in general,  
is responsible for a reduction in the magnitudes of the cross sections. 
Continuing with the relativistic calculations, in \cite{garci2008} the authors considered 
the effects of taking into account 
the width of the $\sigma$ meson in their model which involves the 
coupled $\eta N$-$\pi N$-$\pi \pi N$ subsystems. The $\pi \pi N$ channel is represented by an 
effective $\sigma N$ channel and the authors performed calculations for 
the cases where (i) the $\sigma$ and $\pi$ masses 
are related by $m_{\sigma} = 2 m_{\pi}$ and no width is considered and (ii) the mass and the 
width of the $\sigma$ meson are taken from $\pi \pi$ scattering data. 
The results obtained in \cite{Teresa,garci2008} are shown in Figure 13. 
\begin{figure}[h!]
\centering{
\includegraphics[width=7.5cm,height=6.5cm]{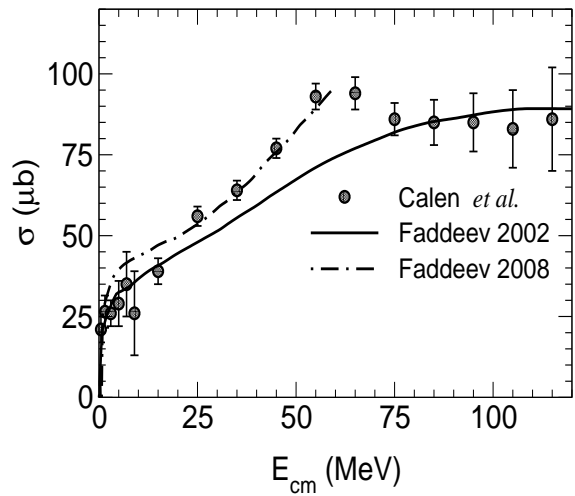}
}
\caption{Cross sections for the reaction $p n \to$ $d \eta$ from a 
non-relativistic Faddeev calculation (solid line) \cite{Teresa} and 
a later calculation by the same authors using a relativistic approach 
(dash dotted line) \cite{garci2008}. 
The latter calculation takes into account the 
finite width of the $\sigma$ meson and was performed at close to threshold 
energies. The data are from Ref. \cite{calen}. 
} 
\end{figure} 
\subsubsection{The $p d \to p d \eta$ reaction}\label{pdeta}
A complete set of data 
covering the excess energy from around threshold to 107 MeV 
exist on the $p d \to p d \eta$ reaction 
\cite{parascatt,hib00,piskor}. 
These data include the invariant mass distribution (integrated 
over other variables) for the $\eta d$, $\eta p$, and $p d$ systems. 
Out of them, like in the case of the $p d \to$ $^3$He $\eta$ reaction, the 
(inclusive) $\eta d$ mass distribution exhibits a large enhancement 
near threshold, hence indicating a strong $\eta d$ attraction. The 
$p d \to p d \eta$ reaction, therefore, like the $p d \to$ $^3$He 
$\eta$ reaction has been studied theoretically in detail. However, 
due to the fact that there are 
three particles in the final state, the incorporation 
of the FSI is much more complicated in this reaction. One theoretical 
effort to understand these data is made in \cite{ulla} where the authors  
essentially explore the role of different reaction mechanisms in 
the production amplitude, but do not include the FSI. In what follows 
we shall discuss Ref. \cite{neelam1} where a detailed study including the 
interaction amongst all the three particles is made. 

The differential cross section for the $p d \to p d \eta$ reaction,
in the center-of-mass system, can be written as \cite{neelam1}
\begin{equation}
\label{sig}
d\sigma\,=\, {m_p^2\,m_d^2 \over 2\,(2\,\pi)^5\,s\,
|\vec{k_p}|}\,d\Omega_{p^\prime}\,\,
|\vec{k_{p^\prime}}|\,\,dM_{\eta\,d}\,\,|\vec{k}_{\eta\,d}|\,\,
d\Omega_{\eta\,d}\,{1\over6}\,\langle\,
|T_{pd \rightarrow pd\,\eta}|^2\,\rangle\,.
\end{equation}

The $T$-matrix which includes the interaction between the $\eta$
and the deuteron is given by,
\begin{eqnarray}\label{one}
T_{pd \rightarrow pd\,\eta} = \langle \, \psi_{\eta d}
(\vec{k}_{\eta d}), \,\vec{k}_{p^\prime};
\,m_{p^\prime},\,m_{d^\prime}\,|\,t_{p d \rightarrow p d 
\eta} \,|\,\vec{k}_p,\,\vec{k}_d;\,m_p,\,m_d\, 
\rangle \, \, \, \, \\ \nonumber
=\langle \, \vec{k}_{\eta d}
(\vec{k}_{\eta d}), \,\vec{k}_{p^\prime};
\,m_{p^\prime},\,m_{d^\prime}\,|\,t_{p d \rightarrow p d 
\eta} \,|\,\vec{k}_p,\,\vec{k}_d;\,m_p,\,m_d\, 
\rangle \\ \nonumber
+ \sum_{m_2^{\prime}} \int \, {d\vec{q} \over (2 \pi)^3} 
{\langle \vec{k}_{\eta d} ; m_{d^{\prime}} | t_{\eta d} | \vec{q}; 
m_{2^{\prime}} \rangle \over E(k_{\eta d}) - E(q) + i \epsilon} 
\langle \vec{q}, \vec{k}_p; m_{2^{\prime}}, m_{p^{\prime}}| 
\,t_{p d \rightarrow p d \eta} \,|\,\vec{k}_p,\,\vec{k}_d);\,m_p,\,m_d\, \rangle  
\end{eqnarray}
where $m_p, \,m_d,\,
m_{p^\prime},$ and $m_{d^\prime}$ 
are the proton and the deuteron spin projections 
respectively. The second line in the above equation comes after replacing 
for $\psi_{\eta d}(\vec{k}_{\eta d})$ using the Lippmann Schwinger equation. 
The $\eta$ d FSI is contained in the $\eta$ d elastic scattering t-matrix 
which is given as \cite{ourtimedelay},  
\begin{equation}
t_{\eta d} (\vec{k}^{\prime}, \vec{k}; z) = 
\langle \vec{k}^{\prime}; \psi_0|t^0(z)|\vec{k};\psi_0\rangle \, +\, 
\epsilon\, \int {d\vec{k}^{\prime\prime} \over (2\pi)^3} 
\, {\langle \vec{k}^{\prime}; \psi_0|t^0(z)|\vec{k}^{\prime \prime}; \psi_0 
\rangle \over (z - {k^{\prime \prime} \over 2 \mu})(z - \epsilon -{ 
k^{\prime \prime}\over 2 \mu })}\, 
\end{equation}
within the finite rank approximation (FRA). The FRA implies that the 
intermediate nucleus in the off shell scattering remains in the ground state. 
$t^0(z)$ is an auxiliary $t$-matrix which is expressed in terms of the 
elementary $\eta N$ t-matrix (see Ref. \cite{ourtimedelay}). 

Another prescription used in \cite{neelam1} involves a factorization
of the half-off-shell $\eta d$ t-matrix into an on-shell part expressed using 
the effective range expansion and off-shell form factors.
Thus \cite{neelam1}, 
\begin{equation}
t_{\eta d} (\vec{k}^{\prime}, \vec{k}; z(k_0))\,= \, g(k,k_0) \, 
{F_{\eta d} (z(k_0))\over (2\pi)^2 \mu_{\eta d}}\, g(k^{\prime}, k_0)\, ,
\end{equation}
with 
$$F_{\eta d}(k) = \biggl ( {1\over a} + {1\over 2} r_0\, k^2\, +\, s\, k^4
\, -\, ik\, \biggr )^{-1}\, .$$
The off shell form factor is written in terms of the deuteron wave function 
as 
$g(k^{\prime}, k_0) = \int d^3r j_0(rk^{\prime}/2) \phi_d^2(r) j_0(rk_0/2)$. 
The effective range parameters $a$, $r_0$ and $s$ are taken from the 
relativistic Faddeev calculation in \cite{garci1}. 

The $p d$ FSI in \cite{neelam1} is included using the Watson-Migdal approach 
\cite{watsonbook, fsibook}, by multiplying the $T$-matrix in (\ref{one}) 
with the Coulomb interaction modified 
inverse Jost function, [J(p)]$^{-1}$. The interaction between 
the $\eta$-meson and the proton in the final state, to a certain 
extent, is contained implicitly in the calculations because in 
the two-step reaction model the $\pi^+ N \to \eta N$ vertex is 
described by a $T$-matrix. The production matrix for the 
$p d \to$ $p d \eta$ reaction in the two-step model is given by  
\begin{eqnarray}\label{bornpde}
< |T_{pd \rightarrow pd\,\eta}| >={3\over2}i \int {d\vec{k}_\pi \over 
(2\pi)^3} \sum_{int\,m's} <p n \,|\,d>
\,<\pi\,d |T_{pp\, \rightarrow\, \pi\, d}| p\,p>
\\ \nonumber
 \times{1\over (k_\pi^2-m_\pi^2+i\epsilon)}
\, <\eta\,p \,|\,T_{\pi N \rightarrow \eta p}\,| \pi\,N>
\end{eqnarray}
where $\vec{k}_\pi$ is the momentum of the exchanged pion and details of the 
dependence of the integrand on it can be found in \cite{neelam1}.
This $T$-matrix is evaluated using the Paris parametrization 
\cite{paris} for the deuteron wave function. The $p d$ Jost function is 
included for both, the spin-doublet and spin-quadruplet states of the
$p d$ system. The expression and details of the Jost function are given
in \cite{neelam1}. 

\begin{figure}[h!]
\centering
\includegraphics[width=8cm]{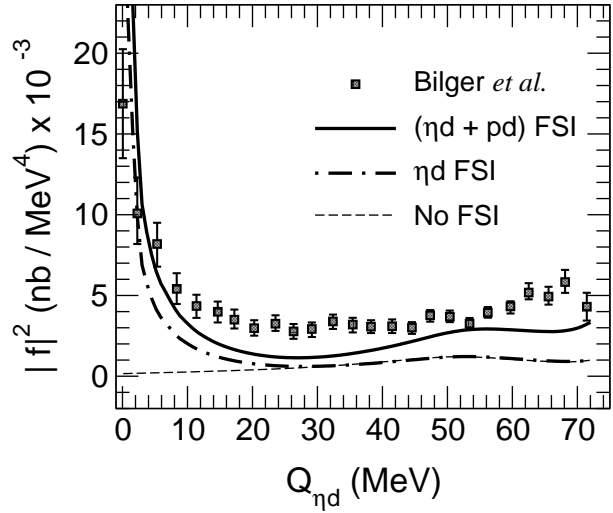}
\caption{Effects of the $\eta d$ and $p d$ FSI on the squared amplitude 
of the $p d \to p d \eta$ reaction 
at a beam energy of 1032 MeV. The $\eta d$ FSI is given by
the factorized prescription using the parameter set corresponding to
a$_{\eta N}$ = 1.07+$i$0.26 fm from Ref. \cite{garci1}. 
$Q_{\eta d} (= M_{\eta d} -m_{\eta} -m_d)$ is the excess energy of the 
$\eta d$ system with $M_{\eta d}$ being the $\eta d$ invariant mass. 
The data are from Ref. \cite{parascatt} and theoretical results from 
\cite{neelam1}.}
\label{pdetafsi}
\end{figure} 
The full squared amplitude along with the data is shown in Figure 
\ref{pdetafsi} including only the $\eta d$ FSI (shown by dot-dashed
line) and the $\eta d$ and $p d$ FSI (shown by the solid line) as a function 
of the excess energy. The $\eta d$ FSI is included within the 
factorized $\eta$ d t-matrix approach. Two results emerge from here:
(1) the enhancement of the observed squared production amplitude 
near threshold is fully described by the FSI and 
(2) while the effect of the $\eta d$ FSI is limited to the excess
energy near threshold, the $p d$ FSI persists over the whole energy
range.

\begin{figure}[h!]
\centering
\includegraphics[width=9cm]{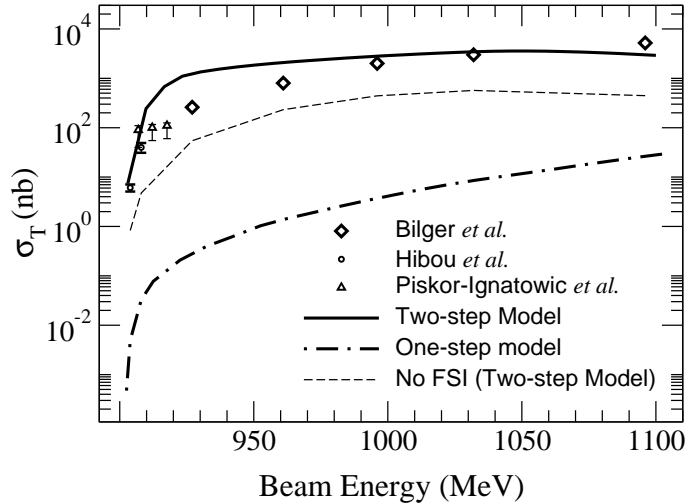}
\caption{The total cross sections for the $p d \to p d \eta$ reaction.
The data from Ref. \cite{parascatt,hib00,piskor} are shown by rhombuses, 
circles and triangles respectively. Theoretical results are from 
\cite{neelam1}. 
}
\label{pdetatot}
\end{figure}
The results for the total cross sections are shown in 
Figure \ref{pdetatot}, where the dashed line represents the plane wave
results while the solid line corresponds to the calculated cross 
sections including both the $\eta d$ and $p d$ FSI. The $\eta d$ FSI is 
evaluated using the factorized prescription for the parameter set 
corresponding to a$_{\eta N}$ = 1.07+$i$0.26 fm taken from the 
relativistic Faddeev calculation of Ref. \cite{garci1}. 
It can be seen that a general agreement with 
the measured cross section is achieved after both the $\eta d$ and 
$p d$ FSI are included. 

As discussed in Section 4.1, the contribution of the one-step model is 
found to be suppressed near the threshold of the meson producing 
reactions because of the large momentum transfer. 
For completeness, the total cross sections calculated using the one-step 
model for the $p d \to p d \eta$ reaction are shown by the dot-dashed 
line in Figure \ref{pdetatot}. The details of this calculation can be 
found in \cite{neelam1}. 

\subsubsection{Cluster model approach to the $p \,^6$Li $\to$ $\eta$ 
$^7$Be reaction}\label{p6li_sec}

As seen in the earlier sections, the measured cross sections near 
the threshold of eta production in $p n$ and $p d$ collisions show large 
enhancements, which are shown to arise from the strong final
state interaction between the $\eta$ meson and $^3$He or deuteron.
This strong FSI is also shown to give rise to quasi-bound eta states
in these nuclei. This observation naturally raises the curiosity 
if such a strong eta-nucleus FSI also exists in heavier nuclei,
indicating thereby the possibility for the 
existence of quasi-bound states in them. 
The first measurement to explore this possibility 
was done in 1993 \cite{scomp} 
by the Turin Group for the  $p \,^6$Li $\to$ $\eta$ $^7$Be reaction
at the beam energy of 683 MeV (excess energy of 19 MeV). 
The differential cross section was found to be 4.6$\pm$3.8 nb/sr 
around 20 degrees. 

A theoretical analysis of this reaction was first performed in 
\cite{khalili} within a cluster model. With the $^6$Li and $^7$Be nuclei 
considered to be d-$\alpha$ and $^3$He-$\alpha$ clusters, the 
$^6$Li ($p$, $\eta$) $^7$Be reaction proceeds via the intermediate 
$d$ ($p$, $\eta$) $^3$He reaction. Performing a simple calculation where 
the cross section for the $^6$Li ($p$, $\eta$) $^7$Be reaction appears as 
\begin{equation}
\biggl ( {d \sigma \over d \Omega} \biggr )_{^6{\rm Li} 
(p, \eta) ^7 {\rm Be}} = ({\rm kinematic} \, \, {\rm factors})\, 
\times \, |F_L(Q)|^2\, \biggl ( {d \sigma \over d \Omega} \biggr )_{{\rm d} 
(p, \eta) ^3{\rm He}}\, ,
\end{equation}
the authors found a good agreement with data for a certain choice of the 
nuclear form factor $F_L(Q)$. The final state interaction (FSI) between the 
$\eta$ meson and the nucleus was not included and the need for improving 
the simple estimate was mentioned.

\begin{figure}[h!]
\centering
\includegraphics[width=12cm]{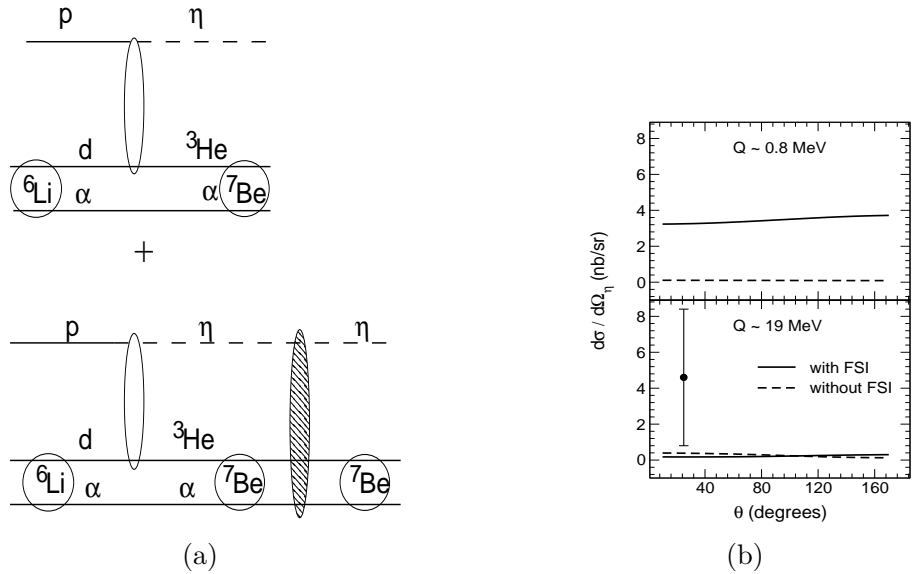}
\caption{The $p \,^6$Li $\to$ $\eta$ $^7$Be reaction: (a) Schematics of the 
cluster model approach. (b) Angular distributions for different excess 
energies, Q. The data point is by the Turin group \cite{scomp}. The 
calculations \cite{neelam7Be} 
are shown for the ground state of the $^7$Be nucleus.}
\label{p6li}
\end{figure}

A more detailed and complete theoretical study of the  p $^6$Li 
$\to$ $\eta$ $^7$Be reaction has been done recently \cite{neelam7Be}. 
This work also uses the cluster model description of the target and the 
recoiling nuclei (see Figure \ref{p6li}(a)). 
Since the reaction is modelled to 
proceed via the $p d$[$\alpha$] $\to$ $^3$He[$\alpha$] $\eta$ reaction 
with the $\alpha$ remaining a spectator, the $\eta$ 
production is governed by the $T$-matrix for the $p d \to$ $^3$He 
$\eta$ reaction. Following the work presented in Section 4.2 this 
$T$-matrix is constructed microscopically 
using the two-step model. 
Neglecting the effect of Fermi 
motion on the $p d \to$ $^3$He $\eta$ production amplitude, 
the $T$-matrix for the $p \,^6$Li $\to$ $\eta$ $^7$Be reaction is 
written as
\begin{eqnarray}
\langle | T_{p ^6Li \to \eta ^7Be}| \rangle&=&i^{(L+1)}\,\sqrt{4\pi}
\,\sum_{M \mu}\,Y^*_{LM}(\hat{Q})\,\,F_L(Q)
\\
\nonumber
&&\hspace{1cm}\times\,\langle J, m_7^\prime| 1/2,\mu, L, M \rangle\,
\,\langle |T_{p d \to \eta ^3He}|\rangle \, ,
\end{eqnarray}
where the transition form factor for $^6$Li $\to$ $^7$Be, $F_L(Q)$, is 
expressed as
\begin{equation}
F_L(Q)\,=\,\int_0^{\infty} r^2\,dr\,\,\Psi_l^{*7}(r)\,\,j_L(Qr)\,\,
\Psi_0^6(r)
\end{equation} 
with the momentum transfer $\displaystyle{\vec{Q}\,=\,{4\over7}\vec{k}
_{\eta}\,-\,{2\over3}\vec{k}_p}$. $\Psi_0^6(r)$ and $\Psi_l^{*7}(r)$
are the radial wave functions for the relative motion of the clusters in 
$^6$Li and $^7$Be respectively 
generated using a Wood-Saxon potential \cite{wood}.

The FSI between the $\eta$ meson and $^7$Be is taken into account as 
discussed before by writing the wave function $\psi_{\eta ^7{\rm Be}}$ 
in the final state using the Lippmann Schwinger equation with an elastic 
$\eta$-$^7$Be transition matrix. The latter is evaluated assuming the 
$\eta$-$^7$Be interaction to be a three body $\eta$-$^3$He-$^4$He problem. 
Thus the $\eta$-$^7$Be t-matrix is written as,
\begin{equation}
T_{\eta ^7{\rm Be}}(\vec{k}^{\prime}, \vec{k}, z)  = 
\int d^3x\, |\Psi_L^7(x)|^2 \, (T_1 (\vec{k}^{\prime}, \vec{k}, 
a_1\vec{x}, z) + T_2 (\vec{k}^{\prime}, \vec{k}, a_2\vec{x}, z) )\, ,
\end{equation}
where $T_1$ and $T_2$ are the medium modified t-matrices for the off-shell 
$\eta$ scattering on the bound $^3$He and $^4$He. $\vec{x}$ is the relative 
coordinate between $^3$He and $^4$He and $\vec{r}_1 = a_1 \vec{x}$ and 
$\vec{r}_2 = a_2 \vec{x}$ are the coordinates of the 2 nuclei with respect 
to the mass-7 center of mass system (with $a_1$ =  4/7 and $a_2$ = -3/7). 
$\Psi_L^7$ represents the inter cluster wave function with angular momentum $L$.
$T_1$ and $T_2$ are themselves written using a Faddeev type decomposition. 
As a result, the $\eta$ meson produced in the final state can undergo 
multiple off-shell scatterings on the $^3$He and $^4$He nuclei until it 
finally emerges as an on-shell meson. 

Figure \ref{p6li} shows the angular distributions of the 
eta meson for different excess energies along with the reaction 
diagram. The angular distributions are seen to remain isotropic at all 
energies considered and the FSI is found to enhance the cross sections near threshold. 

\begin{figure}[h!]
\centering
\includegraphics[width=9cm]{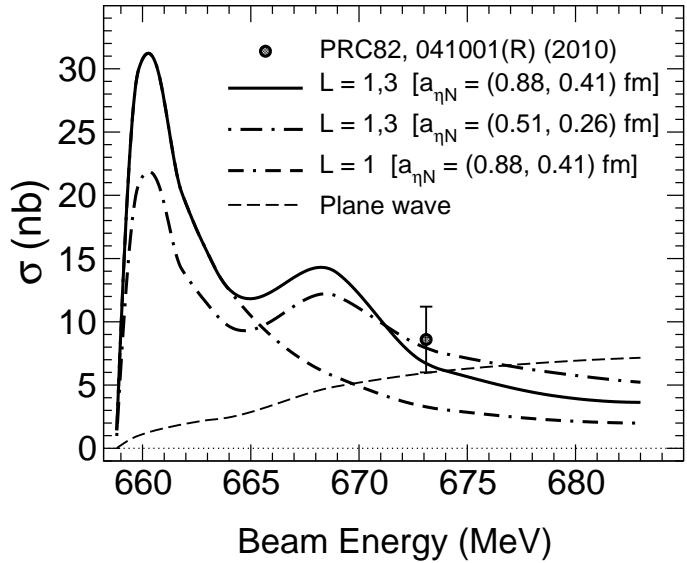}
\caption{Total cross sections for the $p \,^6$Li $\to$ $\eta$
$^7$Be reaction. The solid, dash-dotted and dot-double dashed lines are 
obtained with calculations \cite{neelam7Be} 
including the $\eta \, ^7$Be FSI. The data 
point was measured for the ground state of $^7$Be.}
\label{p6liinclu}
\end{figure} 
Figure \ref{p6liinclu} shows the total cross sections as a
function of the beam energy 
summed over all possible low lying
bound states of the $^7$Be nucleus ($L =1, 3$ included) 
as well as for the case where 
$L=1$. The solid and the dot-dashed lines
in the figure correspond to the results including $\eta$-$^7$Be 
FSI for two different inputs of the $\eta N$ interaction with scattering 
lengths, a$_{\eta}N$ = (0.88, 0.41) fm \cite{fix} and (0.51, 0.26) fm 
\cite{etaN2}. The result with only $L=1$ contribution is shown by the 
double dash-dotted line. 
The figure also shows the results without FSI between 
$\eta$ and $^7$Be, represented by the dashed lines. As in case of 
A=2-4 nuclei, here also the FSI leads to a large enhancement in the 
cross sections up to about 15 MeV excess energy. Two bumps are seen 
in the inclusive distribution. They arise due to different binding energies 
of the different states of $^7$Be, leading to different threshold 
energy for them (for details see Ref. \cite{neelam7Be}).
The data point is from a recent measurement of the total cross 
section made by the COSY
collaboration \cite{gemcosy} for the ground state of $^7$Be. 

This work also determines the scattering length for the 
$\eta$-$^7$Be $T$-matrix describing the FSI in the calculations
for several values of the $\eta N$ scattering length which enter as
an input in the calculations. This is done to explore the possible 
existence of any $\eta$-$^7$Be mesic state. There seems some indication
for such states for large values of a$_{\eta N}$, however it would
indeed be premature to conclude anything about it. It may be better
to perform a time delay analysis or a K-matrix analysis for the
above $\eta$-$^7$Be $T$-matrix.


\section{Summary}
The $\eta$ meson due to its strong attractive interaction with a nucleon has turned out 
to be the most promising candidate for the exploration of exotic states of mesons and nuclei. 
A survey of the theoretical and experimental efforts which triggered and kept the search for 
$\eta$-mesic nuclei alive for the past twenty five years has been performed in the present work. 
In what follows, we summarize some global and specific observations which emerge 
from this survey. 
\begin{itemize}
\item [(i)] A good estimate of the strength of the $\eta$-nucleon ($\eta N$) interaction is crucial for the 
interpretation of the data on $\eta$ meson production on nuclei and the theoretical prediction 
of $\eta$-mesic nuclei. With the possibility of obtaining $\eta$-nucleon elastic scattering data 
being ruled out due to the nonavailability of $\eta$ beams, the $\eta N$ interaction strength is 
quantified in terms of a scattering length, a$_{\eta N}$, determined from 
analyses of $\eta$ production in pion, photon and proton induced reactions. Phenomenological and 
theoretical studies incorporating meson-baryon coupled channels and the relevant intermediate baryon 
resonances obtain a wide range of scattering lengths with 
0.18 $\le$ $\Re$e a$_{\eta N}$ $\le$ 1.03 fm 
and 0.16 $\le$ $\Im$m a$_{\eta N}$ $\le$ 0.49 fm. 
Such a large spread exists despite the fact that the calculations have been 
done in the best possible way using different models.

A closer look at the various 
values of a$_{\eta N}$ obtained in literature, 
however, demonstrates a curious fact. 
Most of the theoretical calculations find 
small values of the real part of a$_{\eta N}$. Though the phenomenological determinations in general 
find both small and big values, one of the most elaborate calculation involving nine baryon resonances 
and the $\pi N$, $\eta N$, $\pi$ $\Delta$, $\rho N$ and $\sigma N$ coupled channels finds 
a$_{\eta N}$ = 0.3 + $i$ 0.18 fm which is very close to the very first prediction [1] 
of a$_{\eta N}$ = 0.28 + $i$ 0.19 fm. A smaller $\eta N$ scattering length would favour light $\eta$ mesic 
nuclei [40] and heavy $\eta$ mesic nuclei with lower binding energies [3,19]. 
\item[(ii)] The experimental searches for the existence of $\eta$ mesic nuclei can be broadly divided into 
two categories; one searching for a direct signal such as a peak in the missing mass spectra and the other 
hinting toward the existence of these states through a large enhancement of the $\eta$ production 
cross sections near threshold. Whereas the former data, due to the 
difficulty in 
performing such experiments are scarce and not conclusive, the latter depends 
on theoretical models for interpretation. The 
theoretical model itself depends on two inputs, namely, 
the reaction mechanism used and the treatment of the final state interaction (FSI) between the $\eta$ meson 
and nuclei. Many of the $\eta$ production cross sections are reproduced well with 
values of a$_{\eta N}$ much larger than those mentioned at the end of 
(i) above. To reach a definite 
conclusion one would have to be sure that both the treatment of FSI as well 
as the reaction mechanism involved bear little uncertainties. 
The FSI between the $\eta$ mesons and nuclei is well determined if one uses 
few body equations rather than resorting to approximate methods for 
incorporating the FSI. The 
uncertainty in the FSI then could arise only from the possible inaccuracy of 
the $\eta N$ transition matrix.

The reaction mechanism at threshold seems to be well understood within models 
based on the two step process.
However, in Chapter 4 we saw that at energies away from threshold where 
the isotropic angular distributions of $\eta$ mesons become forward 
peaked, limitations do exist in the understanding of the reaction mechanisms. 
It could be that at higher energies, there exists an interference rather than 
just a single mechanism which explains the entire 
set of data. Further investigations filling up the missing 
elements in the already existing extensive works is timely. 

\item[(iii)] 
The few body treatment of the FSI shows that the large enhancement 
in the eta-production amplitude near threshold arises mainly due to 
off-shell scattering of the eta-meson in the final state. 
This means that at the production vertex, the $\eta$ meson could 
in principle be produced off-shell, 
undergo multiple elastic scatterings from the nucleus 
and then get converted to an on-shell $\eta$ due to its interaction with 
the nucleus. 
It is important to realize this aspect because many of the eta-meson 
production works in literature consider only 
the on-shell scattering of the eta meson in the FSI. 
Such calculations may necessitate using larger values for 
a$_{\eta N}$ to reproduce the data 
and consequently may lead to unreliability in the conclusion 
about the possible existence of eta-nucleus bound states.

\item[(iv)] The direct signal searches mentioned in (ii) were carried out using protons, pions and photons 
incident on various light nuclei. Some experiments did 
see the signals indicating the existence of 
an $\eta$ mesic state but the results were not reconfirmed. The expectations from 
the future lie in the experiments planned at the J-PARC, MAMI and 
COSY facilities. The K1.8BR beamline at J-PARC for example will be used to perform 
the recoilless production and spectroscopy of $\eta$ mesic nuclei 
using the ($\pi^-$, n) reaction and targets such as $^7$Li and $^{12}$C 
\cite{itahash}. In Tables 1 and 2 of the present review we have collected 
the pole values of all possible unstable states of $\eta$ mesons and nuclei. 
There exists the prediction for quasibound 
eta-mesic $^6$He states from a QCD based 
quark-meson coupling approach \cite{opticalthomas1}. 
Quasibound states of $^{12}$C and $^{16}$O are predicted using different 
approaches \cite{opticalhaider, opticaloset1, opticaloset2, opticalthomas1}. 
However, the pole values from the different approaches differ a lot. 
The predicted range of values can be taken as a guide for the experimental 
searches and the experimental finding of some eta-mesic states could in turn 
confirm the validity of one or more of the theoretical approaches used.

\end{itemize}

\end{document}